\magnification=\magstep1
\input amstex
\documentstyle{amsppt}
\loadeusm
\TagsOnRight
\NoBlackBoxes
\def\tr{\text{tr}}
\def\J{J_{\lambda}}
\def\pd#1#2{\frac{\partial{#1}}{\partial{#2}}}

\topmatter
\title Factorization and the Dressing Method
 for the Gel'fand-Dikii Hierarchy \endtitle 
\rightheadtext{Gel'fand-Dikii Hierarchy} 
\author D.H. Sattinger 
\& J.S. Szmigielski \endauthor
\affil The University of Minnesota  and the University of Kansas\endaffil
\leftheadtext{ D.H. Sattinger \& J.S. Szmigielski}
\footnote[]{This research was supported
in part by 
N.S.F. Grants  DMS-8901607 and DMS-9123844 } 

%8-6-92
\abstract  The isospectral flows
of an  $n^{th}$ order
linear scalar differential operator $L$ under the
hypothesis that it possess a  
 Baker-Akhiezer function were originally investigated by Segal
and Wilson from the point of view of infinite dimensional
Grassmanians, and the reduction of the KP hierarchy to
the Gel'fand-Dikii hierarchy. The associated first order systems and their formal asymptotic
solutions have a rich Lie algebraic structure which was investigated by Drinfeld and
Sokolov. We investigate the matrix Riemann-Hilbert factorizations
 for these systems, and show that  
different factorizations lead respectively to the
 potential, modified, and ordinary Gel'fand-Dikii flows. Lie algebra
decompositions (the Adler-Kostant-Symes method) are obtained for the modified and potential flows. For
$n>3$ the appropriate factorization for the Gel'fand-Dikii flows is not
a group factorization, as would be expected; yet a modification
 of the dressing method still works.
 A direct proof, based on a Fredholm determinant associated
with the factorization problem, is given
that the potentials are meromorphic in $x$ and in
the time variables. Potentials with Baker-Akhiezer
functions include the multisoliton and rational solutions,
as well as potentials in the scattering class with compactly supported
scattering data. The latter are dense in the scattering class.   \endabstract

\endtopmatter
   
\head 1. The Gel'fand-Dikii hierarchies \endhead
  
Gel'fand and Dikii [GD] constructed a hierarchy of isospectral  flows of
the $n^{th}$ order scalar differential operator
$$
L = \sum^n_{j=0} u_j(x)D^{n-j};\quad D = {1\over i}{d\over dx};\quad u_0 =
1,\quad u_1 = 0 .
$$
where $u_j = u_j(x),~j > 2 ,$ are elements of the Schwartz
class $\Cal S(\Bbb R)$.  
The flows are given by
$$
\dot L = [L^{k/n}_+,L],\quad k=1,2,\dots
\eqno(1.1)
$$
where $k\neq 0$ mod $n$, and $L^{k/n}_+$ denotes
the differential part of $L^{k/n}$
considered as a pseudodifferential operator. 
We shall refer to the coefficients
$u_j$ of $L$ as the potential.
The coefficients of $L^{k/n}_+$ are obtained by solving
formal recursion relations and are
universal differential polynomials in the $\{u_j\}.$

The forward and inverse scattering problem for 
$L$ has been investigated
for potentials in $\Cal S$, and
the Gel'fand-Dikii flows for
potentials in this class have been constructed by
Beals [B], and Beals, Deift, and Tomei [BDT]. Action-angle
variables for flows in $\Cal S$
were constructed
in terms of the scattering data by Beals and Sattinger [BS1].                  
Segal and Wilson [SW] investigated these flows under the ansatz that $L$ possess a wave function of Baker-Akhiezer type:
 $w(x,z)\exp\{ixz\}$,
where $w$ is analytic at $z=\infty$ and takes the value 1 there. From now on,
we denote this class of potentials by ${\Cal B}$. 
They analysed the flows from the point of view of infinite dimensional
Grassmanians, and the reduction of the KP hierarchy to
the Gel'fand-Dikii hierarchy. 

The $n^{th}$ order scalar equations can be converted to $n\times n$ first
order systems. These have a rich Lie algebraic structure (cf. Drinfeld and
Sokolov [DS]), just as the Dirac equation rather than the Klein Gordon equation exhibits the internal symmetries of the electron. All reductions are gauge equivalent, but besides the
Gel'fand-Dikki flows, one finds others, such as the modified and potential
Gel'fand-Dikii flows. We analyze the 
matrix Riemann-Hilbert problems corresponding to the first order systems. For potentials in
$\Cal B$, the formal asymptotic series in [DS] are actually
convergent.

Given a Baker-Akhiezer function, it is easily seen that $L$ has
a  basis of such wave functions. On the other hand, there is always a basis of
wave functions which are entire in $z$. Denoting the 
Wronskian of this basis by $\phi_+$ and that of the basis of
Baker-Akhiezer functions by $\phi_-$ we form
the quotient $g=\phi_-^{-1}\phi_+$. The matrices $\phi_{\pm}$ can be chosen        in such a way that $g$ is a function only of $\lambda=z^n.$  Conversely,
appropriate Riemann-Hilbert factorizations of a given a matrix $g(\lambda)$ 
together with a standard argument
(the so-called ``dressing method"), yields a potential in ${\Cal B}.$
Thus, while we cannot 
precisely characterize the class of potentials $\Cal B$, we  
nevertheless can construct them by solving a 
factorization problem. 

All conversions of the $n^{th}$ order equation
to a first order system are gauge equivalent, and the entries
of $L$  are differential polynomials in
the entries of the potential $q$ [DS]. The canonical conversion (by Wronskians) leads to
the Gel'fand-Dikii flows themselves. The modified Gel'fand-Dikii flows
are obtained by first factoring $L$ as a product of $n$ first order
operators, and constructing a system of AKNS type. We investigate in particular the standard and
  modified Gel'fand-Dikii flows together with a third type we call the ``potential" flows. All three are obtained by factoring
exactly the same matrix $g$, but taking the factors in different infinite
dimensional submanifolds.
For the potential and modified flows, 
that submanifold is a subgroup and the components of the connections are obtained by Lie (or loop) algebra decompositions ( \`a la Adler-Kostant-Symes).
For the Gel'fand-Dikii flows themselves the submanifold  is not a group for $n>3$. A Lie algebra decomposition
is not available, but the dressing argument need only be modified slightly.
  
Schilling [Sch] obtained a Lie algebra decomposition for the first order
system of equations for
$\phi_j=L^{(j-1)/n}_+v,\ j=1,\dots,n$, but did not consider the corresponding factorization problem and flows.

The hypothesis that 
$L$ possess a Baker-Akhiezer function is
a very strong one, as observed in [SW]. If $q$ is rational in $x$ the 
wave functions are in general multiple valued in the punctured
neighborhoods of the poles of $q$; but if $L$ has a Baker-Akhiezer
function then the wave functions have trivial monodromy (see \S 2.3).
Moreover, if $q$ is {\it a priori} defined only on some interval
$I$ on the real line, and $L$ has a Baker-Akhiezer function
for $x\in I$, then $q$ has a meromorphic extension to the entire
complex plane (Corollary 3.1.7).

In the case $n=2$, assuming, say, that $(1+|x|)u_2\in L_1( \Bbb R)$,
the assumption implies that the reflection
coefficient is compactly supported. In particular
if $w$ is rational in $z$, then the reflection coefficient vanishes,
and we have a so-called ``reflectionless potential."
Properly interpreted, the same holds for general n,
so that in a sense the ansatz is a generalization of the notion of reflectionless
potential. Segal and Wilson showed that the class $\Cal B$ includes the multi-soliton
solutions, the rational solutions, and in general the ``algebro-geometric"
solutions of the KdV equation. 
We show in this paper that the class $\Cal B$ is actually somewhat
larger, and in fact that $\Cal S_0 \cap
\Cal B$ is dense in $\Cal S_0$ (Theorem 2.1.7). ($\Cal S_0$ is the
class of potentials for which the scattering transform is well-defined
in [BDT]).

We show, by arguments based on the Fredholm determinant of an operator
of trace class associated with the 
factorization problem, that the potentials in $\Cal B$ are meromorphic
in $x$ and in the time variables under the Gel'fand-Dikii flows. Such results 
were proved by Segal and Wilson [SW] using the theory of the $\tau$
function for the Kadomtsev-Petviashvili (KP) hierarchy, 
together with the reduction of the KP hierarchy to
the Gel'fand-Dikii flows. Our proof is similar in spirit but
requires considerably less machinery; furthermore, it
can be extended to other systems in the Drinfeld-Sokolov hierarchy, as well as  the $n\times n$
AKNS flows. 

The non-vanishing of the Fredholm determinant for the system
guarantees both a left and right factorization; whereas the
non-vanishing of the $\tau$ function guarantees only a one-sided
factorization.

The factorization technique used here provides an efficient computational
algorithm for constructing the rational 
and multi-soliton solutions.
We hope to take this up in a future paper.

The restricted class $\Cal B$ is distinct from the potentials for which the scattering
transform of $L$ is well-defined. The ``scattering
class" of potentials is invariant under perturbations by functions 
in the Schwartz class $\Cal S$, whereas $\Cal B$ is
 not.  On the other hand, $\Cal B$ includes the rational
solutions, which, due to their slow decay, or to possible poles on the real 
axis, are not in the scattering class.

For generic potentials in the scattering class, the wave functions are meromorphic
in sectors of the complex plane of angle $\pi/n$. The scattering class leads
to a Riemann-Hilbert factorization
problem on a set of rays $\Sigma$ running from the origin to infinity, [BDT]
while potentials in $\Cal B$ are described by a factorization problem 
on a circle centered at the origin. 
This factorization problem is considerably simpler than that associated with the inverse
scattering problem; moreover, it is closely tied to the infinite dimensional Lie Group/
Lie algebra structure associated with the first order systems
[DS].

In \S2 we develop the forward problem for potentials in $\Cal B$ by showing  
that the ansatz leads to a matrix factorization problem on a circle
centered at the origin in the complex $\lambda$ plane.

In \S3 we consider the inverse problem, formulated as a Riemann-Hilbert
problem. The parameters $x$ and $t$ appear
analytically in the factorization problem, and the solutions of the factorization
problem are shown to be meromorphic in the parameters.  
The dressing method is used to construct a flat connection from
the solutions of the factorization problem and show that the resulting 
connection potentials are meromorphic in $x$ and $t$.  
Similar considerations arise in the proof
of the Painlev\'e property for solutions of isomonodromy deformation problems 
[Mal] (cf. also Miwa, who proved the Painlev\'e property using
the $\tau$ function.)

In \S4  we discuss the infinite dimensional Lie 
structure of the factorization problem, using gradings of the algebra
introduced by Drinfeld and Sokolov. In \S5 we prove that the three factorizations lead to the
standard, modified, and potential Gel'fand-Dikii flows.

This work was carried out during the second author's visit to the IMA
and the School of Mathematics
in the summer of 1991.  He would like to thank the IMA and the University
of Minnesota for their hospitality. Both authors acknowledge
 valuable discussions with R. Beals, J. Dorfmeister,
and I. Herbst concerning various aspects of scattering theory,
factorization theory, and the Gel'fand-Dikii hierarchies.  A number of
techniques used in this paper came out of joint work between
the first author and Richard Beals on the Gel'fand-Dikii flows and
isomonodromy deformations. 

We also benefitted from comments by
the referee and H. Flaschka.

\head 2. The forward problem \endhead
   
\subhead 2.1 Reduction to a first order system \endsubhead
 The scattering theory
 of $L$ proceeds by converting the $n^{th}$ order
scalar equation
$Lf = \lambda f$ to a first order system.  
Let $\phi_j=D^{j-1}f$
for $j=1,\dots , n$. Then the column vector $\phi =
(\phi_1,\phi_2,\dots,\phi_n)^\dag$ satisfies the first order system
$$
(D-J_{\lambda} -q)\phi= 0,    \tag2.1.1
$$
where
$$
J_{\lambda}=\left( \matrix 0 & 1 & 0 &\hdots & 0\\
0 & 0 & 1 &\hdots & 0\\
& &\hdots & & &\\ 
0 & 0 & 0 &\hdots & 1\\
\lambda & 0 & 0 &\hdots & 0\endmatrix \right),
\quad\quad q = - \left( \matrix 0 & 0 &\hdots & 0 & 0\\
0 & 0 &\hdots & 0 & 0\\
& & &\hdots & &\\
0 & 0 &\hdots & 0 & 0\\
u_n & u_{n-1} & \hdots & u_2 & 0\endmatrix \right).
$$

The linear space of all such $q$ 
constitutes a special class of potentials, which we call
the Gel'fand-Dikii potentials. For the rest of this section,
$q$ will denote a potential in this class. In general, we do not
assume the entries of $q$ are in the Schwartz class; but we say that $q\in  
\Cal S(\Bbb R)$ if each entry of $q$ belongs to 
$\Cal S.$
            
Matrix solutions of (2.1.1) are necessarily Wronskians, the entries
of the first row being scalar solutions of  $Lf=\lambda f$. 

Let $f(x,z)$ be a  Baker-Akhiezer function for $L$,  $\alpha_j$, 
$1\le j \le n$, the $n^{th}$ roots of unity, and let $f_j(x,z)=
f(x,\alpha_j z)$. The Wronskian matrix
$W=||D^{k-1}f_j||$, $1\le k,j\le n$,
satisfies (2.1.1). From the asymptotic behavior of $f$ we conclude
$$
\align
W\sim & ||D^{k-1}e^{ix\alpha_j z}||=\Lambda_z e^{ixzJ_{\alpha}}
=d(z)e^{ixzJ_{\alpha}}\Lambda_{\alpha} \\
  z\rightarrow &\infty\\
\endalign
$$                                       
where
$$
J_{\alpha}=\text {diag}(\alpha_1,\dots,\alpha_n), \qquad
d(z)=\text {diag}(1,z,\dots,z^{n-1}), \qquad  \Lambda_z =d(z)\Lambda_{\alpha},
$$  
and
$$
\Lambda_{\alpha}=\left( \matrix 1 & 1 & 1 &\hdots & 1\\
\alpha_1 & \alpha_2 &  &\hdots & \alpha_n\\
\alpha_1^2 & &\hdots & &\alpha_n^2 \\ 
 &  & \hdots & &\\
\alpha_1^{n-1} &  &  &\hdots & \alpha_n^{n-1} \endmatrix \right) . 
$$
The matrix function $\psi=d^{-1}W\Lambda_{\alpha}^{-1}$ satisfies 
$$
D\psi=(zJ+q(z))\psi \qquad \lim_{z\rightarrow \infty}
\psi e^{-ixzJ}=1 \tag2.1.2 
$$
where $q(z)=d^{-1}qd$, $d^{-1}J_{\lambda}d=zJ$, and
$$
J=\left( \matrix 0 & 1 & 0 &\hdots & 0\\
0 & 0 & 1 &\hdots & 0\\
& &\hdots & & &\\ 
0 & 0 & 0 &\hdots & 1\\
1 & 0 & 0 &\hdots & 0\endmatrix \right) . 
$$ 
  
This proves:
\proclaim{2.1.3 Theorem}  $L$ has a Baker-Akhiezer wave function iff
(2.1.2) has a fundamental solution of the form
$\psi=m(x,z)e^{ixzJ}$ where $m(x,z)$ is analytic at $z=\infty$
and tends to $I$ there.
\endproclaim  

Since $\text{tr}(zJ+q)=0$ it follows that $\det \psi$, and hence
$\det m$ is constant in $x$. Since $m\rightarrow I$ as $z\rightarrow \infty$,
$\det m=1$ at infinity. We may therefore
renormalize it so that
$\det m$ is identically 1 in a neighborhood of infinity. 
From now on we shall refer to $m$ as 
a Baker-Akhiezer function for (2.1.2).

We now discuss, briefly, the scattering data of $L$ when $q\in \Cal S$
[BDT].     
Let $\Sigma = \{\xi: \text {Re}\ ix\xi(\alpha_k-\alpha_j) = 0,
\ j,k=1,\dots,\ n\}$, and denote the $2n$ sectors which complement $\Sigma$ by $\Omega_{\nu}, \ \nu=1,2,\dots, 2n.$
Solutions of (2.1.2) are constructed in each $\Omega_{\nu}$
for which the corresponding $m_{\nu}$ satisfies
$$
Dm=[zJ,m]+q(z)m, \tag2.1.4 
$$
and the ``radiation conditions"
$$
\lim_{x\rightarrow -\infty}m(x,z)=I, \qquad \sup_x |m(x,z)|<+\infty,  \tag2.1.5
$$
where $| \cdot |$ denotes the sum of the absolute values of the entries of
the matrix $m$. The wave function $m_{\nu}$ is uniquely determined,
except possibly at a bounded discrete set $Z$,
where it has poles. The points of $Z$ may cluster only along the rays
of $\Sigma$. Away from those cluster 
points, $m_{\nu}$ is continuous up to the boundary 
of $\Omega_{\nu}$. For a generic set of potentials,
denoted by $\Cal S_0,$ 
$Z$ consists of a finite set of points.  
For these $q$ the boundary values of $m$ exist
everywhere on $\Sigma \backslash \{0\}.$ 

The scattering data consists
of the multiplicative jumps (continuous component)
of $m$ across the rays of $\Sigma$, 
together with its  principal part (discrete component) at points in $Z$.
The continuous component of the
scattering data is algebraically 
related, in the case $n=2$, to the reflection
coefficient. The multi-soliton potentials of 
$L=D^2+u$
are examples of potentials in $\Cal B \cap \Cal S_0$ with a trivial 
multiplicative jump.
These are typically referred to as ``reflectionless potentials".
We show below that for potentials in $\Cal B \cap \Cal S_0$ the
multiplicative jumps are compactly supported. Thus $\Cal B$ 
 is an extension of
the class of potentials with compactly supported scattering 
data, and in particular contains the reflectionless potentials.

Let ${\Cal J}$ denote the kernel of ad $J$, and let
$\Cal O$ denote a neighborhood of infinity in the complex
plane.

\proclaim {2.1.6 Lemma} Let $\Cal O$ be a neighborhood of
infinity in the complex plane, let  $m_1$ and $m_2$ be two solutions
of (2.1.4) which are analytic in  $\Cal O \cap \Omega_{\nu}$,
bounded for $x \in \Bbb R$
and, for all real $x$,
tend to $I$ as $z\rightarrow \infty$ in $\Omega_{\nu}$ . Then
$m_1=m_2\delta_{\nu} (z)$, where $\delta_{\nu}$ takes its values in $\Cal J$,
is analytic in $\Cal O \cap \Omega_{\nu}$, and tends to $I$ as $z$
tends to infinity along rays in the interior of $\Omega_{\nu}$. 

If $m_1$ and $m_2$ satisfy (2.1.4) for $z\in \Cal O$,
and
take the value $I$ at infinity, then $m_1=m_2\delta(z)$ in 
$\Cal O$
where $\delta(z)$ takes its values in $\Cal J$, is analytic in 
$\Cal O$, and takes the value $I$ at infinity.
\endproclaim                                            

\demo{Proof} The matrix $w=m_1^{-1}m_2$ satisfies $Dw=[zJ,w],$ 
hence $w=e^{ixzJ}\delta (z)e^{-ixzJ}$ for some 
matrix valued function $\delta$ which is independent of $x$
and 
analytic in 
$\Cal O \cap \Omega_{\nu}$. Since $w$ is bounded
for $-\infty<x<\infty$ and tends to 
$I$ as $z$ tends to infinity, $\delta$ must 
have the required properties. This proves the first statement.

In the second case, $w$ is analytic at infinity, so has a Laurent 
expansion $\Sigma_jw_j(x)z^{-j}$ where the coefficients $w_j$
satisfy the recursion relations $[J,w_{j+1}]=dw_j/dx$, with $w_0=I.$
It follows that $[J,w_1 ]=0$, so that $w_1 \in \Cal J$. Since
$\text{ad}\, J$ is semi-simple, its kernel and range intersect
only at zero. Hence
$[J,w_2]=dw_1/dx$ and $w_1\in \Cal J$ imply that $w_1$ is independent of
$x$ and $w_2\in \Cal J$. By induction, all the coefficients
$w_j$ are independent of $x$ and lie in $\Cal J$. \qed
\enddemo 

We are now ready to prove

\proclaim {2.1.7 Theorem} In the following, let $q\in \Cal S$.
\roster
\item"i." If the continuous component of the scattering data is trivial,
and $Z$ is finite, 
then $L$ has a Baker-Akhiezer function for 
which $m$ is rational in $z$
and is bounded for all real $x$ for regular values of $z$. 
\item"ii." Conversely if $L$ has a Baker Akhiezer function
for which $m$
is rational in $z$ and bounded for all real $x$
for regular values of $z$,
then the continuous component of the scattering
data is trivial.
\item"iii." If the scattering data of $q$ is compactly supported, then
$q\in \Cal B$; and conversely, if $L$ has a Baker-Akhiezer function
for which $m$ is bounded in $x$ for regular values of $z$
then the scattering data is compactly supported. 
\item"iv." $\Cal S_0 \cap \Cal B$ is dense in $\Cal S_0$.
\endroster 
\endproclaim
 
\demo{Proof} To prove i., define a sectionally
meromorphic function $m$ by $m(x,z)=m_{\nu}(x,z)$ for
$z\in \Omega_{\nu}$, 
where  $m_{\nu}$ are the wave functions 
constructed in the forward scattering problem. 
Since $m$ has no jumps across $\Sigma\backslash 0$
and $Z$ is a finite set
$m$ is meromorphic in $\Bbb C \backslash 0$. 
(It takes the value $I$ at infinity.)
By Theorem 8.15 in [BDT], $z^{n-1}d(z)m(x,z)d^{-1}(z)$
has a finite limit at $z=0$. Consequently, $m$ is a rational function of 
$z$ which is bounded in $x$ for regular values of $z$.

Conversely, let $m$ and $m_{\nu}$ denote the
fundamental solutions of (2.1.4)
obtained respectively from the Baker-Akhiezer function and 
the forward scattering problem.
By the first statement in Lemma 2.1.6, $m=m_{\nu}\delta_{\nu}(z)$
for $z\in \Omega_{\nu}$, 
where $\delta_{\nu}$ takes its values in ${\Cal J}$ 
and is  analytic in the intersection of
$\Omega_{\nu}$ with a neighborhood of infinity.
Since
$m_{\nu}$ is continuous up to the boundary of $\Omega_{\nu}$,
$$
m_{\nu}^{-1}m_{\nu+1}(x,\xi)=\delta_{\nu}\delta_{\nu+1}^{-1}(\xi), 
\qquad \xi \in \Sigma_{\nu}, 
$$
where $\Sigma_{\nu}=\Omega_{\nu} \cap \Omega_{\nu +1}$.
Since ad $J$ is a derivation, $s_{\nu}(\xi)=
\delta_{\nu} \delta_{\nu+1}^{-1}\in {\Cal J}$. 

We wish to show that
each $s_{\nu}$ is the identity matrix.
We can just as well work in a representation in which $J$ is diagonal:
hence $J=$diag $(\alpha_1,\ \dots,\ \alpha_n)$,
where $\alpha_j$ denote the roots of unity.
In that case, $s_{\nu}$ is also diagonal, and 
the above equation may be written as
$$
m_{\nu+1}(x,\xi)=m_{\nu}(x,\xi)s_{\nu}(\xi),
$$
or, since $s_{\nu}$ commutes with $J$,
$$
e^{ix\xi J}m_{\nu+1}e^{-ix\xi J}=
e^{ix\xi J}m_{\nu}e^{-ix\xi J}s_{\nu}(\xi)
\tag2.1.8
$$

For $\xi\in\Sigma$, let $\Pi_\xi$ denote the projection
$$
\Pi_\xi e_{jk} = \cases e_{jk} &\text{if}\ Re\ i\xi (\alpha_j-\alpha_k) =
0\cr
0\ &\text{otherwise},\endcases
$$
where $e_{jk}$ denotes the matrix with a 1 in the $jk$ place and zeroes elsewhere. 
We prove below that
$$
 \lim_{x\to -\infty} \Pi_{\xi}e^{ix\xi J}
m_{\kappa}e^{-ix\xi J}  =I+v_{\kappa},
\qquad \kappa=\nu,\ \nu+1, \tag2.1.9
$$
where $v_{\nu}$, $v_{\nu+1}$ lie in the range of $\Pi_{\xi}$ and 
are respectively strictly upper and lower triangular 
in an appropriate ordering of the roots $\alpha_j$.
From the definition of $\Pi_{\xi}$, we see that
$\Pi_{\xi} as = (\Pi_{\xi} a)s$ for a diagonal matrix $s$. Applying
$\Pi_{\xi}$ to (2.1.8) and taking the limit as $x\rightarrow -\infty$,
we obtain
$$
I+v_{\nu+1}= (I+v_{\nu})s_{\nu}.
$$
By the structure of $v_{\nu}$ and $v_{\nu+1}$
we must have $s_{\nu}=I$ and 
$v_{\nu}=v_{\nu+1}=0$.

We establish (2.1.9) by considering
an integral equation for the boundary values of the wave functions. 
For $z\in \Omega_{\nu}\backslash Z$, $m_{\nu}$  
satisfies the 
Fredholm integral equation [B]
$$ 
\align
m_{\nu}(x,z)=I+\int_{-\infty}^x(\Pi_z^-&+\Pi_0)e^{iz (x-y)J}
(qm_{\nu})e^{-iz (x-y)J}dy\\
-\int_x^{\infty}\Pi_z^+&e^{iz (x-y)J}(qm_{\nu})e^{-iz (x-y)J}dy,
\endalign
$$ 
where $\Pi_0$ projects onto the diagonal matrices and
$\Pi_z^{\pm}$ denote the projections 
$$
\Pi_z^{\pm} e_{jk} = \cases e_{jk} & \text{if}\ \pm Re\ iz (\alpha_j-\alpha_k) >
0,\cr
0\ &\text{if}\ \pm Re\ iz (\alpha_j-\alpha_k) <0.\cr
\endcases
$$ 
An ordering of the $\alpha_j$ may
be chosen so that, for $z\in \Omega_{\nu}$,
$\Pi_z^{\pm}$ are the projections 
$\Cal {U}_{\pm}$ onto the strictly
upper and lower triangular matrices. 
Letting $z$ tend to $\xi\in\Sigma_{\nu}$ from $\Omega_{\nu}$,
we obtain the following integral equation for the 
boundary values of $m_{\nu}:$
$$ 
\align
m_{\nu}(x,\xi)=I+&\int_{-\infty}^x (\Cal U_- + \Pi_0)
e^{i\xi (x-y)J}
(qm_{\nu})e^{-i\xi (x-y)J}dy \tag2.1.10 \\
&-\int_x^{\infty}\Cal U_+e^{i\xi (x-y)J}(qm_{\nu})
e^{-i\xi (x-y)J}dy.
\endalign
$$  
Conjugating (2.1.10) by $e^{-ix\xi J}$, applying  
$\Pi_{\xi}$, and taking the limit as $x\rightarrow-\infty$, we obtain 
$$
\lim_{x\rightarrow-\infty}\Pi_{\xi}e^{-ix\xi J}m_{\nu}e^{ix\xi J}=
I-\int_{-\infty}^{\infty}\Pi_{\xi}{\Cal U}_+
e^{-i\xi yJ}(qm_{\nu})e^{i\xi yJ}dy=I+v_{\nu}.
$$

For $z\in\Omega_{\nu+1}$,
$\Pi_z^+$ no longer projects onto the
upper triangular matrices; however,
$\Pi_{\xi}\Pi_z^+= \Pi_{\xi}\Cal U_-$.
Hence, repeating the previous arguments, we obtain
$$
\lim_{x\rightarrow-\infty}\Pi_{\xi}e^{-ix\xi J}m_{\nu+1}
e^{ix\xi J}=
I-\int_{-\infty}^{\infty}\Pi_{\xi}\Cal U_-e^{-i\xi yJ}
(qm_{\nu +1})e^{i\xi yJ}dy=I+v_{\nu+1}.
$$
This establishes (2.1.9), hence ii.

If the scattering data is trivial outside some compact set,
then there is some open neighborhood $\Cal O$ of infinity on which
$m_{\nu}=m_{\nu +1}$ in $\Sigma_{\nu} \cap \Cal O$, and we obtain
a Baker-Akhiezer function by setting $m=m_{\nu}$ for $z\in \Omega_{\nu}
\cap \Cal O.$ 
Hence $q\in \Cal B,$ and iii. is established.

The result in iv) is a consequence of the fact
that potentials with compactly supported scattering data are
dense in $\Cal S_0$ in the Schwartz topology. This fact is 
stated without proof in [BDT] (cf. p. 154).
A rough sketch of the proof runs like this. Let $q\in \Cal S_0$, and
denote its scattering data by $v$. (We also include the
discrete data in $v$.) On $\Sigma$ each entry of
$v-I$ belongs to $\Cal S(\Sigma)$ ([BDT], Theorem 13.1).
In particular, each entry of $v-I$ is rapidly decreasing
at infinty and has an asymptotic expansion in positive powers
of $z$ as $z\rightarrow 0$ along each ray $\Sigma_{\nu}$ of $\Sigma$.
We approximate $v$ by smooth data $v_n$ such that
 $v_n=v$ in $|z| \le n$, and
$v_n-I$ vanishes identically in the region $|z|\ge n+1$.
Since the inverse transform is defined in a sufficiently small neighborhood of
$v$, it is defined for each $v_n$ for sufficiently large $n$;
and the corresponding potentials $q_n$ belong to $\Cal B$. The  inverse 
transform, as constructed in [BDT], is the composition of
two continuous processes: a rational approximation of the scattering
data; and the inversion of a Fredholm operator close to the identity.
Hence the inverse transform is (locally) continous from $L_2(\Sigma)$ 
to $L_2(\Bbb R)$.  Using the estimates in [BDT], one could also prove,
with some effort, that the 
corresponding potentials $q_n$ converge to $q$
in the topology of $\Cal S$.\qed
\enddemo

For potentials in $\Cal S$ the $m_{\nu}$ are 
uniquely determined by their normalization
as $x \rightarrow-\infty$. No such unique normalization is possible
for the Baker-Akhiezer wave functions;
but we do obtain:

\proclaim{2.1.11 Theorem} If $q\in \Cal B$ then a
fundamental solution of (2.1.2) can be constructed 
which satisfies the symmetry condition 
$\psi(x,\alpha z)
=d^{-1}(\alpha)\psi(x,z)d(\alpha)$, whre $\alpha$ is a primitive $n^{th}$
root of unity.
\endproclaim
          
\proclaim{Remark} This symmetry condition does not uniquely determine
the wave function. \endproclaim

\demo{Proof} First note that (2.1.4) is invariant under right
multiplication by  holomorphic $\delta (z)\in \Cal J$ which tend to $I$
as $z$ tends to $\infty$. 
These matrices have the representation
$$
\delta(z)=\sum_{j=0}^{n-1}\delta_j(z)J^j, \qquad \delta_j(z)\rightarrow
\delta_{j0} \ \text {as}\ \ z\rightarrow \infty.
$$        

Since $d^{-1}(\alpha)zJd(\alpha)=\alpha zJ$ and $d^{-1}(\alpha)q(z)d(\alpha)=
q(\alpha z)$, (2.1.4) is also satisfied by 
$d(\alpha)m(x,\alpha z)d^{-1}(\alpha)$. By the second statement of
Lemma 2.1.6 there is a 
$\delta(z)$ with values in $\Cal J$ such that 
$$
d(\alpha)m(x,\alpha z)d(\alpha)^{-1}=m(x,z)\delta (z).
$$
Let $q$ be regular at some point $x_0$, and let $a(z)=m(x_0,z)$. Then 
$$
a(\alpha z)=d^{-1}(\alpha)a(z)\delta (z)d(\alpha).   \tag2.1.12
$$
The function $m$ is uniquely determined by its values at $x_0$. Moreover,
$m\mu$ is also a solution of (2.1.4) for any holomorphic matrix
$\mu(z)$ with values in $\Cal J$. For the proof of Theorem 2.1.11 
it therefore suffices to find a matrix $\mu(z)$
with values in ${\Cal J}$ such that 
$$
d(\alpha)a(\alpha z)\mu( \alpha z)d^{-1}(\alpha)=a(z)\mu (z);  
$$
for it then follows that $m(x,z)\mu$ satisfies the required symmetry
condition. Usi;ng (2.1.12) we may write this condition as 
$$
\delta(z)=\mu (z)d(\alpha)\mu^{-1}(\alpha z)d^{-1}(\alpha)   \tag2.1.13
$$
where
$$
\delta,  \mu \in \Cal J, \qquad \delta(\infty)=I, \qquad \det \delta =1.
$$
We represent $\delta$ and $\mu$ in the form 
$$
\delta(z)=\exp\{\sum_{j=1}^{n-1}\eta_j(z)J^j\}, \qquad \mu(z)=
\exp\{\sum_{j=1}^{n-1}\theta_j(z)J^j\}.
$$
Iterating (2.1.13) $n$ times, we obtain 
$$
I=\delta(z)d(\alpha)\delta(\alpha z)\dots\delta(\alpha^{n-1}z)d(\alpha).
$$
Bringing each of the
$n$ matrices $d(\alpha)$ all the way to the left in the above
expression (and noting that $Jd(\alpha)=d(\alpha) \alpha J$),
we find
$$
I=\exp\{\sum_{j=1}^{n-1} \sum_{k=0}^{n-1}(\eta_j(\alpha^k z)\alpha^{-k})
J^j\},
$$    
and therefore
$$
\sum_{k=0}^{n-1}\eta _j(\alpha^k z)\alpha^{-k} = 0, \quad j=1,\dots,n-1  
$$ 
These equations represent a set of constraints on the $\eta_j$,
which are given functions.
From (2.1.13) we find by similar arguments,
$$
\eta_j(z)=\theta_j(z)-\alpha^{-j}\theta_j(\alpha z). \tag2.1.14  
$$
We seek a solution of (2.1.14) in the form
$$
\theta_j(z)=\sum_{k=0}^{n-2}v_k\eta_j(\alpha^k z), \quad j=1,\dots,n-1,    
$$
where the coefficients $v_k$ (more precisely $v_{j,k}$)
are to be determined. 
The three preceeding equations reduce to the linear system 
$$
\left( \matrix 1 & 0 & \hdots &  & \alpha^{-j}\\
-\alpha^{-j} & 1 & 0 &\hdots & \alpha^{-(j+1)}\\
0 &-\alpha^{-j} & 1 & 0 & \alpha^{-(j+2)}\\ 
 &\hdots  &  &\hdots & \\
0 &    &\hdots &-\alpha^{-j}& 1+\alpha^{-(j-2)}\endmatrix \right) 
 \left( \matrix v_0 \\ v_1\\.\\.\\ v_{n-2}\endmatrix \right)
= \left( \matrix 1\\0\\.\\.\\0 \endmatrix \right). 
$$   

The coefficient matrix on the left is invertible, as can be
seen by reducing it to row echelon form. Adding the 
first row to the second multiplied by $\alpha^j$ , etc.
one arrives at a matrix in row echelon form whose diagonal is
$$
1,\ \alpha^j, \ \alpha^{2j},\ \dots,-\alpha^{-(j-1)}( 1-\alpha ^{-j}).
$$
The last entry is non-zero, since $\alpha$ is a primitive 
$n^{th}$ root of unity. \qed
\enddemo 
 
\subhead 2.2 Properties of the wave function \endsubhead

We next analyze the structure of the wave function $m$ constructed in theorem
2.1.11. From the symmetry condition
$$
\psi(x,\alpha z)=d^{-1}(\alpha)\psi(x,z)d(\alpha) \tag2.2.1
$$
and the expansion
$$
m(x,z)=\sum_{j=0}^{\infty} \frac {m_j}{z^j}, \qquad m_0=I \tag2.2.2
$$
we deduce 
$$
d^{-1}(\alpha)m_jd(\alpha)=\alpha^{-j}m_j, \quad j\ge 0. \tag2.2.3                  
 $$
We claim that
$$
m_j(x)= E_j(x)J^{-j}, \qquad E_j \ \text {diagonal},
\qquad E_0=I.   \tag2.2.4
$$
First, from the identity  $e_{ij}=e_{ii}J^{j-i}$ we see that any
$A\in M_n(\Bbb C)$, the space of $n\times n$ matrices 
with complex entries, can be represented as
$$
A=\sum_{k=0}^{n-1}E_kJ^k, \qquad  E_k \ \text{ diagonal}.\tag2.2.5
$$
Furthermore,
$$ 
d^{-1}(\alpha)J^kd(\alpha)=(\alpha J)^k,  \qquad k\in{\Bbb Z}.  \tag2.2.6
$$
Expanding $m_j$ in the form (2.2.5) and using 
(2.2.6) we see that $E_k=0$ for $j \ne k \ \text {mod}\  n$. 
This establishes (2.2.4).

From now on we often suppress the $x$ dependence of the
$E_j$.  Further restrictions are placed on them by the
requirement that $m$ satisfy (2.1.4). 
First we note that  
$$
q(z)=\sum_{j=1}^{n-1}Q_{j+1}(zJ)^{-j}, \qquad 
Q_j=\text{diag}\ (0,0,\dots,-u_j). \tag2.2.7  
$$ 
Substituting (2.2.2) into (2.1.4) we obtain a sequence
of recursion relations for the $m_j=E_jJ^{-j}$.  
Straightforward calculations lead to 
$$ 
\align
 (\sigma - I)E_1=&  0,   \tag2.2.8 \\
(\sigma -I)E_{j+1}= &  DE_j -\sum_{k=1}^j Q_{k+1} \sigma^{-k}(E_{j-k}),
\qquad j\ge 1  
\endalign
 $$
where we set $Q_k=0$ for $k> n$ and $\sigma (E_j)=JE_jJ^{-1}$.
We shall need the following simple facts about $\sigma$.
\proclaim{2.2.9 Lemma} Let $\Cal D $ denote the algebra of diagonal $n\times
n$ matrices and let $\Cal D_0$ be the subspace of traceless such   
matrices.
Then
\roster
\item"i)"  $\sigma \in  \text{Aut}\ \Cal D$
\item"ii)" $\sigma ^n=I$ 
\item"iii)"  Range $ (\sigma - I)=\Cal D_0$,\quad  Ker($\sigma - I)=\Bbb C\ I$ 
\item"iv)" On $\Cal D_0$ and for $n>2$,                                          
$$
(\sigma - I)^{-1}=\frac {1}{n} ( \sigma ^2+2\sigma ^3+\dots+(n-2)\sigma ^{n-1}
-I).
$$ 
\endroster  
\endproclaim
\demo{Proof} Items (i-iii) are
immediate from the definition of $\sigma$, while (iv)
follows from the identity
$$
I+ \sigma + \sigma ^2+\dots +\sigma^{n-1}=0\qquad \text {on}\  \Cal D _0.
$$
\enddemo

\proclaim{2.2.10 Theorem} Let $D_1$ denote the matrix diag
$(1,1,\dots,1,1-n)$. Then 
$$ 
\align
E_0 =&I, \qquad   
E_1 = \omega_1I,\\
E_j =&\omega_jI+\sum_{k=1}^{j-1}(D^k\omega_{j-k})(\sigma-I)^{-k}D_1, \qquad j\ge 2.
\endalign
$$
The $\omega_j$ are determined from the coefficients
$u_j$ of $L$ by a sequence of recursion relations 
$$
u_j=-nD\omega_{j-1}+h_j\qquad 2\le j\le n
$$
where $h_2=0$ and $h_j$ is a polynomial in $\omega_1,\ \dots,\ \omega_{j-2}$
and their derivatives for $j>2$. 
\endproclaim
\demo{Proof} By (2.2.8) $E_1$ is a scalar multiple of the identity, so
$E_1=\omega_1I.$ By the second relation in (2.2.8) 
$$
(\sigma-I)E_2=DE_1-Q_2I. 
$$ 
This equation is solvable provided that
$\text {tr}\ (DE_1-Q_2)=nD\omega_1+u_2=0;$
in that case
$$
DE_1-Q_2=D\omega_1I-nD\omega_1e_{nn}=D\omega_1(I-ne_{nn})=
(D\omega_1)D_1,
$$
and 
$$
E_2=\omega_2I+(D\omega_1)(\sigma-I)^{-1}D_1.
$$
Assume the result holds for $j\ge 2$. Since $(\sigma-I)^{-1}$ maps
$\Cal D_0$ to itself, $\tr DE_j=\tr D\omega_jI=nD\omega_j$, hence
the recursion relation (2.2.8) for $E_{j+1}$ is solvable provided
$$
nD\omega_j-\tr\ \sum_{k=1}^jQ_{k+1}\sigma^{-k}(E_{j-k})=0. \tag2.2.11 
$$
It follows from (2.2.11) that
$$
\sum_{k=1}^j Q_{k+1}\sigma^{-k}(E_{j-k})=
e_{nn}\tr \sum_{k=1}^j Q_{k+1}\sigma^{-k}(E_{j-k})=n(D\omega_j)e_{nn},
$$
so that (2.2.8) may be written as
$$
\align
(\sigma-I)E_{j+1} = & D\omega_jI+\sum_{k=1}^{j-1}D^{k+1}\omega_{j-k}
(\sigma-I)^{-k}D_1-\sum_{k=1}^jQ_{k+1}\sigma^{-k}(E_{j-k})\\
= & (D\omega_j)D_1+\sum_{k=1}^{j-1}D^{k+1}\omega_{j-k}(\sigma-I)^{-k}D_1\\ 
= & \sum_{k=1}^jD^k\omega_{j+1-k}(\sigma-I)^{-k+1}D_1.
\endalign
$$ 
Therefore
$$
E_{j+1}=\omega_{j+1}I+\sum_{k=1}^j D^k\omega_{j+1-k}(\sigma -I)^{-k}D_1.
$$                                  
For $2\le j \le n-1$ (2.2.11) implies
$$
\align
nD\omega_j & +\sum_{k=1}^ju_{k+1}\text {Tr}
[e_{nn}\sigma^{-k}(E_{j-k})]\\ 
=nD\omega_j & +u_{j+1}+\sum_{k=1}^{j-1}u_{k+1}\text {Tr}
[e_{nn}\sigma^{-k}(E_{j-k})]=0.
\endalign
$$
This yields the appropriate recursion relation for $\omega_j$
and the induction step is completed.\qed
\enddemo  

Now we are ready to state the main theorem of this section.

\proclaim{2.2.12 Theorem}Let $\tilde m =d(z)md^{-1}(z)$.
Then
\roster
\item"i)"  
$$
\tilde m=\sum_{j=0}^\infty E_jJ_{\lambda}^{-j},
$$
where the $E_j$'s are given in Theorem 2.2.10;

\item"ii)"$\tilde m(x,\alpha z)= \tilde m(x,z)$, hence
$\tilde m=\tilde m(x,\lambda)$;

\item"iii)"  $\tilde m$ is regular at $\lambda=\infty$ and
 $$
\lim_{\lambda \to {\infty}}\tilde m(x,\lambda)
=I+\ell,
$$
where $\ell$ is a strictly lower triangular matrix. 
\item"iv)"  det $\tilde m=1$
\endroster
\endproclaim
\demo{Proof}
Observe that $d(z)zJd^{-1}(z)=J_{\lambda}$ 
and  $E_j$  diagonal imply that
$d(z)E_j(zJ)^{-j}d^{-1}(z)=$ $ E_jJ_{\lambda}^{-j}$. This proves i) and ii)
follows immediately.
By the definition
$$
\tilde m=dmd^{-1}=\sum_{j=0}^\infty E_jJ_{\lambda}^{-j}=
\sum_{j=0}^{n-1}E_jJ_{\lambda}^{-j}+O(|z|^{-n}).
$$
Note however that 
$$
\lim_{z\to \infty}J_\lambda^{-1}=\left( \matrix
0&0&\hdots&0\\1&0&\hdots&0\\
\vdots&\ddots&\hdots&0\\
0&\hdots&1&0\endmatrix \right) \equiv K.
$$
Thus
$$
\tilde m(x,\infty)=\sum_{j=0}^{n-1}E_jK^j=I+\ell
$$
where $\ell$ is strictly lower triangular.
Finally $\det \tilde m=\det m=1$.
\qed
\enddemo
 
\subhead 2.3 Factorization for the Gelfand-Dikii flows.\endsubhead

The existence of a Baker-Akhiezer function for $L$ leads to a 
factorization problem on a circle centered at the origin in the
$\lambda$ plane; specifically:

\proclaim{2.3.1 Theorem} Let $q\in \Cal B$. Then there exist wave
functions $\phi_{\pm}(x,\lambda)$ of (2.1.1) such that $\phi_+$
is entire and $\phi_-=\tilde me^{ixJ_{\lambda}}$, where  $\tilde m(x,\infty)
=I+\ell$, $\ell$ strictly lower triangular. $\phi_{\pm}$
satisfy the factorization problem 
$$
g(\lambda)=\phi_-^{-1}(x,\lambda)\phi_+(x,\lambda),   \tag2.3.2
$$                      
where $g$ is analytic in  $\rho < |\lambda|<+\infty$ for  
some $\rho>0.$ By a suitable rescaling of the potentials,
we may assume that $\rho<1$.
\endproclaim
 
\demo{Proof} The wave function 
$$
\phi_-=d(z)\psi_-d^{-1}(z)=d(z)m_-(x,z)d^{-1}e^{ixJ_{\lambda}}=
\tilde m(x,\lambda )e^{ixJ_{\lambda}}
$$
satisfies (2.1.1) and, by Theorem 2.2.12, has all the 
required properties.  
Since (2.1.1) depends analytically on $\lambda$, 
there is a unique solution 
$\phi_+(x,\lambda)$, which is entire in $\lambda$ and
satisfies the initial value condition $\phi_+(0,\lambda)=I$. (We assume,
for convenience, that $q$ is regular at $x=0$.) Since $\phi_{\pm}$  both
satisfy (2.1.1), the matrix  $\phi_-^{-1}\phi_+$ is independent of $x.$

Given a Baker-Akhiezer function for the potentials $u_j(x)$ 
which is analytic in the exterior of some circle in the
$\lambda$-plane, we  define rescaled potentials by
$u_j^s(x)=s^ju_j(sx),\ s>0.$
If $q^s(x)$ denotes the matrix of rescaled potentials,
then $\phi^s=d(s)\phi(sx,s^{-n}\lambda)d^{-1}(s)$ satisfies
$(D-J_{\lambda}-q^s(x))\phi^s=0$. Replacing the original
wave functions by the rescaled ones, we obtain a factorization
problem on any circle in the $\lambda$ plane we choose. In
particular, we may assume  (2.3.2) holds
for some $\rho<1$.\qed
 \enddemo

The isospectral flows of $L$
are compatibility conditions for the pair of scalar equations
$$
Lf=\lambda f \qquad D_tf=(L^{k/n}_+)f, \quad  D_t=-i\partial / \partial t.
$$ 
By (2.1.1), all $x$-derivatives of $f$
can be written as linear combinations of the first $n-1$ derivatives  
of $f$ with coefficients which are differential polynomials in
the $u_j$.  The equation for the time evolution of $f$ is
therefore equivalent to a first order system of the
form $(D_t-v)W=0$, where $W=||D^{k-1}f_j||$
is the Wronskian introduced in \S2.1. Since the coefficients
of $L^{k/n}_+$ are differential polynomials in the coefficients
of $L$, $v$ also possesses such a structure. When $q=0$ we find
that $v=J_{\lambda}^k$, so the Gel'fand-Dikii flows are
zero curvature conditions for the connection 
$D-J_{\lambda}-q, \  D_t-J_{\lambda}^k-B_{k,n},$
$B_{k,n}$ denoting a matrix valued differential polynomial in 
$u_j$ of degree $k-1$ in $\lambda$.     
 Let $q(x,t)\in \Cal B$ evolve according to
the $(n,k)$ Gel'fand-Dikii flow. The wave functions $\phi_{\pm}$
of Theorem 2.3.1
satisfy 
$$ 
(D-J_{\lambda}-q)\phi_{\pm}=0, \qquad  (D_t-J_{\lambda}^k-B_{k,n})\phi_{\pm}=0.
$$
They may be written as 
$$
\phi_{\pm}=\tilde m_{\pm}(x,t,\lambda)e^{i(xJ_{\lambda}+tJ_{\lambda}^k)},
$$
where $\tilde m_+$ is an entire function of $\lambda$,
and $\tilde m_-$ satisfies the conditions of Theorem 2.1.12.
From Theorem 2.3.1 we have:

\proclaim{2.3.3 Corollary} The functions $\tilde m_{\pm}$ satisfy the factorization problem
$$
e^{i(xJ_{\lambda}+tJ_{\lambda}^k)}g(\lambda)e^{-i(xJ_{\lambda}+tJ_{\lambda}^k)}
=\tilde m_-^{-1} \tilde m_+,
\tag2.3.4
$$ 
where again $g$ is analytic in an exterior domain containing the unit circle,
except possibly at $\lambda=\infty$.
\endproclaim   

Later, in Theorem 4.3.4, we shall see that the factorization
given in (2.3.4) is unique. Using this, we may prove:
\proclaim{ 2.3.4 Theorem} Suppose $q$ is rational in $x$ in a domain
$\Cal U$ and in addition is a Baker-Akhiezer potential. Then 
the monodromy of the wave functions is trivial.
\endproclaim
\demo{Proof} Suppose
$$
e^{xJ_{\lambda}}ge^{-xJ_{\lambda}}=m_-^{-1}m_+
$$
and suppose $\hat m_{\pm}$ denote the
analytic continuations of $m_{\pm}$ around a pole
of $q$. Since the process of analytic continuation
is continuous, we must have
$$
e^{xJ_{\lambda}}ge^{-xJ_{\lambda}}=\hat m_-^{-1}\hat m_+.
$$
By the uniqueness of the decomposition (Theorem 4.3.4 below), we
must have $\hat m_{\pm}=m_{\pm}.$ \qed \enddemo

\head 3. The Inverse Problem \endhead

\subhead 3.1 Factorization on the circle \endsubhead

In \S2 we derived the factorization problem (2.3.4) on the unit circle, 
beginning with  $L$ and its isospectral flows 
for potentials in $\Cal B$. The inverse problem 
consists of solving a factorization problem and constructing the
flat connection from the factors $\tilde m_{\pm}$. We begin with
a brief summary of the solution of matrix factorization problems
on the unit circle. (cf. [GK].)
                          
Let $G$ denote the loop group 
$$
G=\{g:g=g(\lambda)\in M_n(\Bbb C),\
  \det\ g=1,\  g\ \text {analytic in}\ 
\rho <|\lambda|<+\infty\}
$$
together with its subgroups (under standard matrix multiplication)
$$ 
\align
G_+=&\{g\in G:g=\sum_{k\ge 0}g_k\lambda^k \};\\
G_-=&\{g\in G:g=\sum_{k\le 0}g_k\lambda^k, \quad g_0=I \} 
\endalign
$$ 
We always assume $\rho <1$, so that the domain of holomorphy 
of $g$ contains the unit circle.
In practice, we use $g$ to signify either the function $g(\lambda)$ 
or the sequence of Fourier coefficients $\{g_k\}$.
For $g\in G$  
$$
\sum_k|g_k|s^k<+\infty
$$
for all $s>\rho,$ where $|g_k|$ denotes the sum
of the absolute values of the entries of $g_k.$
Therefore for any $s>\rho$, $G$ is a subset of the Hilbert space $H^s$
defined by 
$$
H^s=\{g:||g||^2= \frac{1}{2\pi} \int_0^{2\pi}\text {tr}\ [g(
s e^{i\theta})
(g(s e^{i\theta})^{\ast})]\ d\theta =
\frac{1}{2\pi}\sum_ks^{2k} \text{tr}\ g_kg_k^{\ast}<+\infty\}.
$$ 
We define 
$$
H_+^s=\{ g\in H^s : g_k=0\  \text{for}\ k< 0\}, \quad H_-^s=
\{ g\in H^s: g_k=0\  \text{for}\ k\ge 0\}.
$$

Then $H_{\pm}^s$ are the Hilbert spaces of boundary values of 
matrix-valued functions holomorphic 
in the interior and exterior of the unit 
circle. From now on we sometimes drop the superscript $s$.
Moreover, if $\phi_-\in H_-^s$ for all $s>\rho$ and $\det \phi_-=1$ then 
$\phi_-\in G_-$.

We associate a Fredholm determinant with the
factorization problem. (Segal and Wilson obtain the
$\tau$ function as a determinant associated with a Riemann-Hilbert problem.)
The Fredholm determinant is defined
for operators of the form $I+A$ where $A$
is of trace class on a Hilbert space $H$ [Si].
The operators of trace class  on $H$ form a 
complex Banach space $\eusm I_1$
with  norm   
$$
||A||_1=\text {tr}\sqrt{AA^{\ast}}.
$$ 
The nonlinear functional det $(I+A)$
is well defined and differentiable on  $\eusm I_1$. Under the
identification of $\eusm I_1^{\ast}$ with the space
of bounded linear operators on $H$, its Frechet derivative is
$$
(I+A)^{-1}\det (I+A).
$$ 
The operator
$$
D(A)=-A(I+A)^{-1}\det (I+A)
$$
is the first Fredholm minor, and 
$$
(I+A)^{-1}=I+\frac{D(A)}{\det (I+A)}.
$$
Since $\det \ (I+A)$ is differentiable on the
complex Banach space $ \eusm I_1$ 
it is analytic there, and therefore so is $D(A)$.

\proclaim{Theorem 3.1.1} If $A(x)\in \eusm I_1$
is holomorphic in a complex variable  $x$,
and $\det (I+A(x))\ne 0$ for at least one $x\ne x_0$,
then $(I+A(x))^{-1}$ is meromorphic in $x$,
with poles at the zeroes of det $(I+A(x))$.\endproclaim

In the course of the solution of the factorization problem
we need to consider operators on $H_{\pm}$ of the following type:
$$
C_g^+f=\sum_{k\ge 0}f_kg_{j-k}, \ j\ge 0; \qquad 
C_g^-f=\sum_{k<0}f_kg_{j-k},\ j< 0.  
$$  

\proclaim {3.1.2. Theorem} For $g\in G$,
$C_g^{\pm}$ are Fredholm operators on $H_{\pm}$.
\endproclaim      

\demo{Proof} Let $K_g$ denote the operation
defined on $H$ by $K_gf=fg$, where $g \in G$
and $f \in H$. The equation $K_gf=h$
can be written as the infinite system of equations
$$
\sum_kf_kg_{j-k}=h_j.  \tag3.1.3
$$ 
This system is invertible for $g \in G$ since
$g$ is bounded and
det $g(\lambda)=1$ on the unit circle,
and $K_g$ has a bounded inverse on $H$.

We rewrite (3.1.3) as  
$$  
\align
\sum_{k\ge 0}f_k^+g_{j-k}+\sum_{k>0}f_k^-g_{j+k}=&h_j^+, \qquad j\ge 0, \\ 
 & \tag3.1.4 \\                                 
\sum_{k\ge 0}f_k^+g_{-j-k}+\sum_{k>0}f_k^-g_{k-j}=&h_j^- \qquad j>0
\endalign
$$
where $f_k^+= f_k, \ k\ge 0; \  f_k^-=f_{-k}, \ k>0; etc.$

 Denoting  $\{f^+,f^-\}$ as well by $f$, we introduce
the operators
$$
B_g  f=  \left( \sum_{k=0}^{\infty}f_k^+g_{j-k},\ 
\sum_{k>0} f_k^-g_{k-j} \right), \qquad
T_g f=   \left( \sum_{k>0}f_k^-g_{j+k},\ 
\sum_{k=0}^{\infty} f_k^+g_{-(k+j)} \right).
$$ 
By the decomposition (3.1.4), $K_g=T_g+B_g$,
and $B_g=K_g-T_g=K_g(I-K_g^{-1}T_g).$ The operators $C_g^{\pm}$
are precisely the components of $B_g$.
Since $K_g$ is invertible, it suffices to show that $T_g$ is compact.

In fact, each of its components is of trace class. Consider
for example, the operator on $H_+$ defined by $Tf=\Sigma_kf_kg_{k+j}$.
Now an infinite matrix $A=||a_{jk}||$ is of trace class if
$$
\sum_{j,k \ge 0}|a_{jk}|<+\infty.
$$  
In fact, (cf. [Si]) there is a partial isometry $U$ such that
$|A|=U^{\ast}A$, where $|A|$ is the positive definite operator
$(AA^{\ast})^{1/2}$. Since $||U^{\ast}\psi ||=||\psi ||$, all entries
of $U^{\ast}$ are bounded by one in absolute
value. Consequently
$$
||A||_1=\text{tr}|A|=\sum_{j,k}U^{\ast}_{jk}a_{kj}\le \sum_{j,k}|a_{jk}|.
$$ 
Thus it suffices to show that $\Sigma_{j,k}|g_{j+k}|<+\infty$
where $|g_k|$ denotes the sum of the absolute values of
$g_k$. We have
$$
\sum_{j,k\ge 0}|g_{j+k}|=\sum_{n=0}^{\infty}(n+1)|g_n|;
$$
and the sum on the right converges since $g$ is analytic in
$|\lambda|>1.$  

The second component of $T_g$ is also of
trace class. A similiar argument leads to consideration
of the series
$$
\sum_{n\le 0}|g_n|(1+|n|).
$$
This sum converges since $g$ is analytic for
$\rho<|\lambda|<1$.\qed \enddemo

Since $T_g$ is of trace class, the following is immediate:

\proclaim{3.1.5. Corollary}  The operators $C_g^{\pm}$  are
 invertible iff
det $(I-K_g^{-1}T_g)\ne 0$. If $g$ is holomorphic
in $x$ then $(C_g^{\pm})^{-1}$ are meromorphic in $x$.
\endproclaim

Let
$$
G^t=\{g:g=g(\lambda)\in M_n(\Bbb C),\
  \det\ g=1,\  g\ \text {analytic in}\ 
t <|\lambda |<+\infty\},
$$
and define $H^t,\ H_{\pm}^t$ and $G_-^t$ accordingly.
We are now ready to prove:

\proclaim{3.1.6. Theorem} Let $g\in G^{\rho}$. The factorization problem 
$$
\phi_-g=\phi_+, \qquad g\in G,\quad \phi_{\pm} \in G_{\pm}^t, \quad t>\rho,
$$ 
has a unique solution
provided  $\det(I-K_g^{-1}T_g)\ne 0$. Moreover, if $g$ is analytic in a parameter
$x$, the factors $\phi_{\pm}$ are meromorphic in $x$, provided that
for at least one value of $x$, $\det(I-K_g^{-1}T_g)\ne 0$.
\endproclaim
 
\demo{Proof} Writing
$$
\phi_-=I+\sum_{k<0}\phi_k\lambda^k, \qquad \phi_+=\sum_{k\ge 0}\phi_k\lambda^k,
$$
and substituting these expressions into our factorization problem, we get
$$
\sum_kg_k\lambda^k+\sum_k\left( \sum_{j<0}\phi_jg_{k-j}\right)
\lambda^k=\sum_{k\ge 0}\phi_k\lambda^k.
$$
For $k<0$ we get the system of equations
$$
g_k+\sum_{j<0}\phi_jg_{k-j}=0, \qquad k<0.  \tag3.1.7
$$
By Corollary 3.1.5 this system has a unique solution in $H_-^t$, denoted by
$\phi_-$, provided that $\det (I-K_g^{-1}T_g)$ does not vanish. Setting
$$
\phi_k=g_k+\sum_{j<0}\phi_jg_{k-j}, \qquad k\ge 0, \tag3.1.8
$$
we obtain the required factorization
provided we can show that $\phi_{\pm}$ belong to $G_{\pm}^t$
respectively. Since $\phi_-$ and $g$
are analytic in $|\lambda|>t$, so is $\phi_+=\phi_-g$. The factorization
shows that there are no negative powers of $\lambda$ in the Laurent
expansion for $\phi_+$. Therefore $\phi_+$ is entire. 
 
Note that $\det \phi_+= 
\det \phi_-$ on $|\lambda|=1$ so the sectionally holomorphic
function given by
$$
\cases
        & \det\phi_-, \quad |\lambda |>1\\
        & \det\phi _+, \quad |\lambda |<1 
\endcases
$$
is analytic on the extended complex plane. Therefore it is a constant.
Since $\phi_-(\infty)=I$, it is identically 1,
and det $\phi_{\pm}\equiv 1$. Hence $\phi_+$ is entire and
$\phi_{\pm} \in G_-^t.$ Since $\phi_-$ does not depend on $t$ it belongs
to $G_-^t$ for all $t>\rho$, hence $\phi_-\in G_-$.

If  $g$  depends analytically on $x$, then the  $K^{-1}T_g$ is an
analytic operator valued function of $x$, and the solution
$\gamma$ of (3.1.7) is meromorphic in $x$. If $g$ is entire, then
the poles of $\phi_-$ do not cluster
in the finite complex $x$ plane. \qed
\enddemo

\subhead 3.2 The dressing method \endsubhead
 Once the factorization of $g\in G$
has been obtained, a flat connection is reconstructed from the factors 
by a procedure often referred to as the ``dressing method" 
[ZS].
We first  summarize the method for the AKNS hierarchies with potentials in 
$\Cal B$ since these are somewhat less complicated.

Given $g\in G$,
let $m_{\pm}$ be the solutions of the factorization problem
$$
e^{xzJ+tz^k\mu}g(z)e^{-xzJ-tz^k\mu}=m_-^{-1}m_+.    \tag3.2.1
$$
where $J$ and $\mu$ are diagonal matrices. It is immediate
from (3.2.1) that
$$
D(m_-^{-1}m_+)=z[J,m_-^{-1}m_+],\qquad 
D_t(m_-^{-1}m_+)= z^k[\mu,m_-^{-1}m_+], \tag3.2.2
$$
where for the present $D=d/dx$ and $D_t=d/dt$. 

Denote by $m$ the sectionally
holomorphic function
$$
m=\cases m_+ & |z|<1 ;\\ m_- & |z|>1. \endcases
$$ 
A simple calculation based on (3.2.2) shows that the sectionally
holomorphic functions
$$
(Dm)m^{-1}+m(zJ)m^{-1}, \qquad  (D_tm)m^{-1}+m(z^k\mu )m^{-1}
$$
have no jumps across the unit circle, and are consequently 
entire functions of $z$. Since $m_-$ tends to $I$ as $z$
tends to infinity, their highest order terms in $z$ 
are respectively  $zJ$ and $z^k\mu$; hence, with $m=m_-$,
$$
\align
(Dm)m^{-1}+m(zJ)m^{-1}=& u(x,z)\equiv zJ+q(x,t) \tag3.2.3a \\ 
(D_tm)m^{-1}+m(z^k\mu )m^{-1}=&v_k(x,z) \equiv  z^k\mu +B_k
\tag3.2.3b
\endalign
$$
where $B_k$ denotes a polynomial of degree $k-1$ in $z$.

Letting $(\ )_+$ denote the projection onto the non-negative 
powers of $z$, we obtain
$$
zJ+q=(m_-zJm_-^{-1})_+=zJ+[m_1,J]   \tag3.2.4
$$ 
where $m_1$ is the coefficient of $z^{-1}$ in the expansion
of $m_-$, and 
$$
v_k(x,z)=(m_-z^k\mu m_-^{-1})_+.   \tag3.2.5
$$
\proclaim{3.2.6 Theorem} $v_k$ is a universal polynomial in $q$ 
and its derivatives with zero constant term. Specifically,
$$
v_k=\sum_{j=0}^kF_jz^{k-j}
$$
where the matrix valued functions $F_j$ satisfy the
recursion relations
$$
[J,F_{j+1}]=DF_j-[q,F_j], \quad F_0=\mu.        \tag3.2.7
$$ 

The connection $D-u,\ D_t-v_k$ is flat,
and $q$ satisfies the nonlinear partial differential equation
$$
q_t=[J,F_{k+1}].     \tag3.2.8
$$ 
Solutions of (3.2.8) which are obtained from the factorization 
(3.2.1) are meromorphic in $x$ and $t$.

Likewise, the coefficients of $m\J^k m^{-1}$, where
$me^{x\J}$ satisifes (2.1.1), are differential
polynomials in $q$ with zero constant term.

\endproclaim
\demo{Proof} We sketch a proof of the first statement based on arguments in [BS2]. The wave function
$m_-$ is uniquely determined up to right multiplication by
a diagonal matrix, so $F=m_-z^k\mu m_-^{-1}$ is independent of
the choice of $m_-$. For fixed $x_0$, construct a formal series
$m_-$ by requiring the diagonal entries in $m_j,\ j>0$ to vanish at
$x_0$. It is clear that the coefficients of
$m_-$ are then uniquely determined polynomials in $q$ and its derivatives at $x_0$, so
the same is true of the $F_j$. Since $x_0$ is arbitrary, the result
holds for all $x$. When $q$ is zero, $m_-$ is necessarily diagonal and
$F=z^k \mu$; hence these polynomials have zero constant term. The same argument holds for the Gel'fand-Dikii
case, where $m_-$ satisifies $m_x=[\J,m]+qm$ and 
$F=m\J^k m^{-1}$.

To prove the second statement, rewrite (3.2.3) in
the form $(D-u)m=m(D-zJ), \  (D_t-v)m=m(D_t-z^k\mu).$
Then 
$$
[D-u,D_t-v]m=m[D-zJ,\ D_t-z^k\mu ]=0.
$$
Since $m$ is invertible, $[D-u,\ D_t-v]=0$ and the connection
is flat. The evolution equation (3.2.8) follows
in the standard way from this fact and the recursion
relations (3.2.7).

Since $x$ and $t$ appear analytically in 
(3.2.1), the factors $m_{\pm}$ are meromorphic
in $x$ and $t$ by Theorem 3.1.6. The meromorphic
behavior of $q$ follows from (3.2.4).\qed \enddemo 

The wave function $m$ is said to ``dress" the bare connection
$D-zJ,\ D_t-z^k\mu$ to the ``dressed connection"
$D-u,\ D_t-v$. 

Flat connections for the Gel'fand-Dikii and associated hierarchies
are constructed from solutions of the factorization problem (2.3.4),
but  require a number of algebraic considerations. Namely, one
must decide how one is going to split the constant factors
in the Lie group $G$. By altering the
choice of $G_{\pm}$ one obtains not only the
Gel'fand-Dikii flows, but their associated flows as well,
such as the modified Gel'fand-Dikii flows, etc. Moreover, in the
case of the Gel'fand-Dikii flows themselves, the factor $G_-$
in the Riemann-Hilbert splitting is not a group, so the
dressing method must be modified. This will be done in \S5.1.

The analogs of (3.2.3a,b) are 
$$
\align
(D\tilde m)\tilde m ^{-1}+ \tilde mJ_{\lambda}\tilde m^{-1}=
&J_{\lambda}+q \tag3.2.9a \\
(D_t \tilde m)\tilde m^{-1}+ \tilde mJ_{\lambda}^k\tilde m^{-1}
&=J_{\lambda}^k+B_{n,k}.  \tag3.2.9b
\endalign
$$

\proclaim{3.2.10 Corollary} Let $q$ be defined
for $x\in I \subset \Bbb R$ and suppose $L$ has a Baker-Akhiezer
function for $x\in I$. Then $q$ has a meromorphic extension
to the complex plane.
\endproclaim
\demo{Proof} For $x\in I$ the wave functions satisfy the
factorization problem (2.3.4) (with $t=0$). Since the left side 
is entire in $x$, the factors $m_{\pm}$ have meromorphic
extensions to the complex plane, by Theorem 3.1.6. The
meromorphic extension of $q$ is then given by (3.2.9a).
\qed 
\enddemo
This result was obtained in [SW] using the $\tau$ function;
the method here applies to the AKNS and Drinfeld-Sokolov hierarchies
as well.

\head 4. Algebraic structure \endhead

\subhead 4.1 Loop groups and algebras \endsubhead
 We begin by
reviewing the basic ideas of loop groups and algebras
[K], [PS]. We continue to denote by $G$ the 
loop group introduced in \S3.1. Let $|\cdot|$ denote
a norm on $M_n(C)$. We may assume that the automorphism
$\sigma$ on diagonal matrices $\Cal D$,
which we introduced in  
\S2.2, leaves $|\cdot|$ invariant: $|\sigma (D)|=
|D|.$  

From the identity
$$
\lambda^me_{jk}=e_{jj}J_{\lambda}^{k-j+nm}, 
$$ 
it follows that an equivalent representation of $G$ is 
given by
$$
G=\{g:g=\sum_{k=-\infty}^{\infty}d_kJ_{\lambda}^k;
\ \ \det\ g=1,\ \ d_k\in \Cal D \} 
$$
where it is assumed that the series converges
absolutely and uniformly on $S^1$.  Given
$g=\Sigma_kd_kJ_{\lambda}^k,$ and $h=\Sigma_kh_kJ_{\lambda}^k$
we have
$$
gh=\sum_j \left( \sum_kd_k\sigma^k(h_{j-k}) \right)
J_{\lambda}^j.
$$
Moreover, $G$ is a Banach Lie group with norm 
$$
||g||=\sum_k|d_k|,
$$  
and Banach Lie algebra 
$$
\frak g=\{h:h=\sum_kh_kJ_{\lambda}^k;\ ||h||<+\infty,  
\ \text {tr}\ h_k=0,\ \text {for}\ k=0\ \text {mod}\ n\}. 
$$

An integer valued {\it grading} of $\frak g$ 
is a direct sum decompostion of $\frak g$ into  
subspaces $\frak g_k$ such that
$
[\frak g_j,\ \frak g_k] \subset \frak g_{j+k}.
$ 
In our case, $\frak g$ has two natural
gradings, induced by the two bases:
$$
\text{deg}_1\ (g_k\lambda^k)=k, \qquad \text{deg}_2\ (d_kJ_{\lambda}^k)=k.
$$ 
The homogeneous subspaces are
different for the first and second gradings; let us denote them
by $\frak g_k^1$ and $\frak g_k^2.$ Then:

\proclaim{4.1.1 Lemma} The complex dimension of
$\frak g_k^1$ is $n^2-1$ while
$$
\text{dim}\ \frak g_k^2=\cases n &  \text {if}\ k\ne 0\ \text{ mod}\ n\\
        n-1 & \text{otherwise}. \endcases
$$  
 \endproclaim 

\subhead 4.2 Lie algebra decompositions \endsubhead
Let 
$$
V_n=\bigoplus_{j=-n+1}^{-1}\frak g_j^2=\{a:a=
\sum_{k=-n+1}^{-1}a_kJ_{\lambda}^k, \quad a_k\in \Cal D\}.     
$$
It is easily seen that dim $ V_n=n(n-1)$ and that 
$$
V_n=\Cal U_-\oplus \Cal U_+\lambda^{-1}
$$
where $\Cal U_{\pm}$ are the subspaces of strictly
upper and lower triangular matrices. 

$G$ and $\frak g$ are contained in the
linear space
$$
B =\{a:a=\sum_ka_kJ_{\lambda}^k,\ a_k\in \Cal D \}
        =\{a:a=\sum_kg_k\lambda^k,\ g_k\in M_n (\Bbb C)\}.
$$ 
This space also has gradings relative
to the two bases, as above. 
We define projections $P_1$ and $P_2$ (denoted respectively by
$P_+$ and $P^+$ in [DS]) onto the
subspaces of non-negative degree relative to 
these gradings by 
$$
P_1a=\sum_{k\ge 0}g_k\lambda^k, \qquad P_2a=
\sum_{k\ge 0}a_kJ_{\lambda}^k,  
$$
where
$$
a=\sum_kg_k\lambda^k=\sum_k a_kJ_{\lambda}^k.
$$

We define a third projection on $B$ as follows.
Let $D_1$ be as in Theorem 2.2.10.  

\proclaim{4.2.1 Theorem} There is a projection
$P_3$ whose image coincides with that of  
$P_1$ and whose kernel is given by
$$
\text{ker}\ P_3= \{g:g=\sum_{k>0} e_kJ_{\lambda}^{-k}, \quad
e_k\in \Cal E_k\}
$$
where 
$$
\Cal E_k=\cases \text{span}\  [\ I,\ (\sigma-I)^{-1}D_1,\ \dots \ , 
\ (\sigma -I)^{-(k-1)}D_1], & k\le n-1 ; \\  
    \Cal D & k\ge n. \endcases
$$
\endproclaim

\demo{Proof} We must show that every $a\in B$ can be written
uniquely as $a=h+v$, where $v\in$ Im $P_1$ and $h$
belongs to the subspace ker $P_3$ defined above. We may assume
$a$ is homogeneous with respect to the second grading.
If deg$_2\ a\ge 0$, then $a\in$ Im $P_1$, and $a=v$.
On the other hand, if deg$_2 \ a\le -n$, then 
$a\in$ ker $P_3$ and $a=h$.

That leaves elements in $V_n$.
If deg $a=-j$ then $a=a_jJ_{\lambda}^{-j}$ for
some $a_j\in \Cal D.$ We must show that
$$
a_j=e+f,   \qquad e\in \Cal E_j,\ f\in \Cal F_j  \tag4.2.2
$$ 
where  
$$
\Cal F_j=\{f:f=\text{diag}(f_1,\dots,f_n),\ f_1=f_2=\dots=f_j=0\}.
$$ 

It is clear that dim $\Cal F_j=n-j$. We first prove that
dim  $\Cal E_j=j$ and so it
suffices to prove that $\Cal E_j\cap\Cal F_j=\{0\}$.
Suppose that
$$
e=\omega_0I+\sum_{r=1}^{j-1}\omega_r(\sigma-I)^{-r}D_1=0.
$$
Then
$$
(\sigma-I)^{(j-1)}e=\omega_1(\sigma-I)^{j-2}D_1 +\dots 
+\omega_{j-1}D_1=0.
$$
It is easily checked that $(\sigma-I)^kD_1\in \Cal F_{n-k-1}$
for $k<n$;
and this implies that $\omega_{j-1}=0$.  
Applying
$(\sigma -I)^{j-2}$ to $e$ we find next that $\omega_{j-2}=0$
and so forth. Hence all the coefficients are zero,
and dim $\Cal E_j=j$.

Now let $e$ be as given above and suppose that
$e\in \Cal E_j \cap \Cal F_j$. We repeat the above
argument. Applying $(\sigma-I)^{j-1}$ to $e$ 
we obtain
$$
(\sigma-I)^{(j-1)}e=\omega_1(\sigma-I)^{j-2}D_1 +\dots 
+\omega_{j-1}D_1\in (\sigma-I)
^{j-1}\Cal F_j \subset \Cal F_1.
$$ 
  Hence $\omega_{j-1}=0$. Then we apply
$(\sigma-I)^{j-2}$ to $e$, and find 
that $\omega_{j-2}=0$ and so forth. Hence
$e=0,$ and $ \Cal E_j\cap
\Cal F_j=\{0\}$.\qed
\enddemo

The decomposition (4.2.2)
can be obtained by
multiplying on the right by  $e_{kk}$ with $k<j$
and taking the trace. In that case we obtain the system
of equations
$$
\tr [e_{kk}a]=\omega_0+\sum_{r=1}^{j-1}\omega_r \tr [e_{kk}(\sigma-I)^{-r}D_1],
\quad 1\le k \le j,
\tag4.2.3
$$
for the coefficients $\omega_0,\dots,\omega_{j-1}$.
For future reference we denote the matrix of coefficients
of (4.2.3) by $W$. 

The projection $P_3$ is obtained by splitting $V_n\subset$ ker $P_2$
 into the direct sum of two subspaces. The dimension of
ker $P_3\cap V_n$ is $n(n-1)/2$.

Let us set
$$
\frak g_j^-= \text{ker}\ P_j\cap \frak g,
 \qquad \frak g_j^+= \text{im}\ P_j\cap \frak g, \qquad j=1,2,3.
$$
Then 

\proclaim{4.2.4 Theorem}\roster
\item"i)" $\frak g=\frak g_j^-
\oplus \frak g_j^+$ for $j=1,2,3.$
\item"ii)" $\frak g_1^-, \frak g_2^-,\frak g_1^+,
\frak g_2^+,$ and $\frak g_3^+$ are each 
Lie subalgebras of $\frak g$ for all $n$.
\item "iii)"$\frak g_3^-$ is a subalgebra
of $\frak g$ for $n=2,3$ but not
for $n>3.$ 
\endroster
\endproclaim

\demo{Proof} Statement i) follows from the fact
that each $P_j$ is a projection. That the kernel
and range of $P_1$ are subalgebras is immediate
from the definition.
To prove the corresponding fact for $P_2$,
note first that $J_{\lambda}bJ_{\lambda}^{-1}
=\sigma (b)$ for any diagonal matrix $b$. 
Then, for $a$ and $b$ diagonal,
$$
[aJ_{\lambda}^j,bJ_{\lambda}^k]=(a\sigma^j(b)-b\sigma^k(a))
J_{\lambda}^{j+k}.
$$ 
It follows that the kernel and range of
$P_2$ are subalgebras, since they are the
subspaces of elements spanned respectively by 
negative and positive powers of $J_{\lambda}$.

To prove iii) it suffices to consider the subspace
$$
\frak g_3^-\cap V_n = {\text sp}\{e_kJ_{\lambda}^{-k}, \ k=0,\dots,n-1,
\quad e_k\in \Cal E_k \}
$$
For $n=2$, this subspace is spanned by $J_{\lambda}^{-1}$.
Since $[J_{\lambda}^{-1},\ J_{\lambda}^{-1}]=0$,
it is trivial that $\frak g_3^-$ is a subalgebra. 
For $n=3$ one must also check the commutator
$$
[J_{\lambda}^{-1},(\sigma -I)^{-1}D_1J_{\lambda}^{-2}]=-\sigma^{-1}
(D_1)J_{\lambda}^{-3}.
$$  
The element on the right has degree -3 in the second grading. 
When $n=3$ it automatically belongs to 
$\frak g_3^-$, since in that case 
no further constraint is placed on the
diagonal matrix multiplying $J_{\lambda}^{-3}.$

For $n\ge 4$, however, the 
condition that it belong to $\frak g_3^-$ is
$$
\sigma^{-1}D_1=\alpha I+\beta (\sigma -I)^{-1}D_1 +
\gamma (\sigma-I)^{-2}D_1
$$
for some complex constants $\alpha, \beta, \gamma $.
Operating on this equation by $\sigma (\sigma-I)^2,$
we find
$$
(\sigma^2(1-\beta)+(\beta-2-\gamma)\sigma+1)D_1=0.
$$
It is not hard to see, from the definition
of $D_1$, that these equations are inconsistent.
Thus $\frak g_3^-$ is not a subalgebra when $n\ge 4.$ \qed
\enddemo

\subhead 4.3 Lie group decompositions\endsubhead
 The
above Lie algebra decompositions have their 
counterparts
on the group level. Let
$$
\gather
G_j^+=\{g: g\in G,\ g\in \text{Im}P_j\},             \tag4.3.1  \\
G_j^-=\{g:g\in G,\ g-I\in \text{Ker}P_j\}.             \tag4.3.2
\endgather
$$  
Then: 
$$ 
\gather
G_1^-= \{g:\det g=1,\quad g=\sum_{j\le 0}g_j\lambda^j, \quad g_0=I\}  \\
G_1^+=\{g:\det g=1,\quad g=\sum_{j\ge 0}g_j\lambda^j\},
\endgather
$$  
\vskip .1in
$$
\gather
G_2^-=\{g:\det g=1,\quad g=\sum_{j\le 0}g_jJ^j_{\lambda},
 \quad g_j\in \Cal D,
\quad g_0=I\}\\
G_2^+=\{g:\det g=1,\quad g=\sum_{j\ge 0}g_jJ^j_{\lambda}, \quad g_j\in \Cal D\},
\endgather
$$
and
$$
\gather
G_3^-= \{g:\det g=1,\quad g=\sum_{j\le 0}g_jJ^j_{\lambda},
 \quad g_j\in \Cal E_j,
\quad g_0=I\}\\   
G_3^+=G_1^+
\endgather
$$

\proclaim {4.3.3 Theorem} 
\roster
\item"i)" $G_3^+=G_1^+$.
\item"ii)" $G_2^+$ is a Lie subgroup of $G_1^+$ and
$G_1^-$ is a Lie subgroup of $G_2^-$.
\item"iii)" $G_j^-$ and $G_j^+, \ j=1,2$
are Banach Lie groups with Lie algebras $\frak g_j^-$
and $\frak g_j^+$.
\item "iv)" $G_3^-$ is a group only for $n=2,3$.
\endroster
\endproclaim

This theorem is proved in the same manner as Theorem 4.2.5; the 
demonstration is left to the reader.

We are now ready to state the main theorem of this section:

\proclaim{4.3.4 Theorem} For each of the three projections $P_j$ there
is a an open dense subset $\Omega_j \subset G$ such that
each $g\in \Omega_j$ has a unique factorization
$$
g=(\phi_j^-)^{-1}\phi_j^+, \quad \phi_j^{\pm}\in G_j^{\pm}.
$$
\endproclaim 

\demo{Proof} For $j=1$ this is the well-known Riemann-Hilbert
factorization whose proof is given in Theorem 3.1.6. 
Now suppose $g\in \Omega_1$ admits such a factorization 
with $ \phi^{\pm}\in G_1^{\pm}.$ 
We write
$$
\phi^+(\lambda)=v\hat \phi^+(\lambda), \quad \hat \phi^+(0)=I, \ \ 
v=\phi_+(0).
$$  

Let $SL^{\circ}(n)$ denote the subset of 
$v$ in $SL(n)$ which have the factorization
$v=\ell^{-1}u$ where $\ell =I+strictly\ lower$ and $u$ is
an upper triangular matrix. As is well known,
$SL^{\circ}(n)$ is a dense open subset of $SL(n)$.
When $v\in SL^{\circ}(n)$, $g$ factors as 
$$
g=(\ell \phi^-)^{-1}(u\hat \phi^+),
$$
where $\ell \phi^-\in G_2^-$ and $u\hat \phi^+\in G_2^+$.

The domain $\Omega_2$ consists of all $g\in \Omega_1$ for which
$v=\phi^+(0)\in SL^{\circ}(n)$. Since $\Omega_2$ is the inverse
image of $SL^{\circ}(n)$ under the continuous mapping
$g\rightarrow v$, it is open. To show that $\Omega_2$ is
dense in $\Omega_1$, consider $g\in \Omega_1\backslash 
\Omega_2$. From (3.1.8),
$$
v=g_0+\sum_{j<0}\phi_jg_{-j} .
$$ 
Let $\dot g$ denote a tangent vector at $g$ to a curve in $G$.
For those curves for which $\dot g$ is  
independent of $\lambda$, $\dot v=
\dot g_0$. The set of such tangent vectors is transversal
to $SL(n)\backslash SL^{\circ}(n)$; for example, it contains all strictly 
triangular matrices. So every neighborhood of $g$ contains points in
$\Omega_2.$

The construction of the third factorization is more involved.
We begin by proving: 

\proclaim {4.3.5 Lemma} Given $a$ in $G_1^-$ or $G_2^-$   
there is a unique constant lower triangular matrix $\ell$ with 1's on the 
diagonal such that $\ell a\in G_3^-$.
\endproclaim

\demo{Proof} It suffices to prove the lemma for $a\in G_2^-$,
since $G_1^-$ is a subgroup of $G_2^-$.  Let
$$
a=\sum_{k=0}^{\infty} a_kJ_{\lambda}^{-k}, \quad a_k\in \Cal D,\ a_0=I.
$$
Any constant lower triangular $\ell$ with 1's on the diagonal can be
written as
$$
\ell=\sum_{j=0}^{n-1}f_jJ_{\lambda}^{-j},\quad f_0=I,\ f_j\in \Cal F_j.
$$   
It follows that
$$
\ell a=\sum_{k=0}^{\infty}\left(\sum_{j=0}^kf_j\sigma^{-j}(a_{k-j})\right)
J_{\lambda}^{-k}
$$
where we set $f_j=0$ for $j\ge n.$ Thus $\ell a\in G_3^-$ if and only if
$$
\sum_{j=0}^kf_j\sigma^{-j}(a_{k-j})\in \Cal E_k,\qquad  k=0,\dots,n-1.
$$
Writing these equations out we get
$$ 
\gather
f_1+a_1=e_1 \\
f_2+f_1\sigma^{-1}(a_1)+a_2=e_2 \\ 
                \vdots         \\
f_{n-1}+\sum_{k=1}^{n-2}f_k\sigma^{-k}(a_{n-k-1})+a_{n-1}=e_{n-1} .
\endgather
$$
But these decompositions are precisely those obtained in (4.2.2).
Given $a_1$ the first equation may be solved uniquely for
$f_1$ and $e_1$. Then $f_2$ and $e_2$ are uniquely determined
from $a_1$ and $a_2$ from the second equation, and so forth.\qed
\enddemo

We now return to the proof of Theorem 4.3.4. We set $\Omega_3=\Omega_1$.
Given $g\in \Omega_1$, let $g=\phi_1^-\phi_1^+$. By the preceding lemma
there exists a unique lower triangular matrix $\ell$ with
1's on the diagonal such that $\ell \phi_1^-\in G_3^-$. Note that
left multiplication by a constant matrix of determinant
1 leaves $G_1^+=G_3^+$ invariant; and therefore
$$
g=(\ell \phi_1^-)^{-1}(\ell \phi_1^+), \qquad \ell \phi_1^{\pm}\in G_3^{\pm}
$$
is the desired factorization.

For $j=1,2$,  the uniqueness of the factorizations  
is a simple consequence of the fact that
$G_j^-$ and $G_j^+$ are groups and
$G_j^- \cap G_j^+=\{I \}$. 
For $j=3$, however, $G_3^-$ is not
 a group for $n>3$, and
we have to argue differently.

Let us assume therefore that $g=(\phi_3^-)^{-1}\phi_3^+=(\tilde
\phi_3^-)^{-1}\tilde \phi_3^+.$  Let $\ell$, respectively $\tilde \ell$
denote the values of $\phi_3^-$ and $\tilde \phi_3^-$ at
infinity; $\ell $ and $\tilde \ell$ are lower triangular
matrices with 1's on the diagonal. Writing
$\phi_3^-=\ell v_3$ and $\tilde \phi_3^-=\ell
\tilde v_3$, we have
$$
g=v_3^{-1}(\ell^{-1}\phi_3^+)=\tilde v_3^{-1}
(\tilde \ell^{-1}\tilde \phi_3^+),          \tag4.3.6
$$ 
where now 
$$
v_3,\ \tilde v_3\in G_1^-, \qquad
 \ell^{-1}\phi_3^+,\ \tilde \ell^{-1} \tilde \phi_3^+ \in G_1^+.
$$
The factorization (4.3.6) of $g$ is of type 1, and is therefore 
unique, by what we have already proved, so $v_3 =
\tilde v_3$. By Lemma 4.3.5 there is a unique matrix
$\ell$ such that such that $\ell v_3=\ell \tilde v_3
\in G_3^-$. Thus $\ell=\tilde \ell$, $\phi_3^-=\tilde \phi_3^-$, and $\phi_3^+=
\tilde \phi_3^+$. \qed
\enddemo

\head 5. Nonlinear Flows \endhead
                                            
\subhead 5.1 The Gel'fand-Dikii flows \endsubhead
  In the preceeding section
we introduced three different factorizations of 
elements of $ G$. These three factorizations lead to
the Gel'fand-Dikii, modified and ``potential" Gel'fand-Dikii flows. In this section
we discuss the construction of these different flows from the different
factorizations. We begin with the Gel'fand-Dikii flows themselves. 

\proclaim {5.1.1 Theorem} Let $g$ belong to $G$ and let $m_{\pm}$ 
be the solutions of the factorization problem
$$
e^{i(xJ_{\lambda}+tJ_{\lambda}^k)}g(\lambda)e^{-i(xJ_{\lambda}+tJ_{\lambda}^k)}
=m_-^{-1}(x,t,\lambda)m_+(x,t,\lambda)
\tag5.1.2
$$
where $m_{\pm}\in G_3^{\pm}$.
Then 
$$
(Dm)m^{-1}+mJ_{\lambda}m^{-1}=J_{\lambda}+q, \qquad m=m_{\pm},   \tag5.1.3
$$
where $q=q(x,t)\in \Cal B$ satisfies the Gel'fand-Dikii equations for
$(k,n)$
\endproclaim

\demo{Proof} By the results of \S3 and Theorem 4.3.4
the factors $m_{\pm}$ of
(5.1.2) are meromorphic functions of $x$ for fixed $t$ and
of $t$ for fixed $x$.  By the arguments of \S3.2,
$(Dm)m^{-1}+mJ_{\lambda}m^{-1}$, where $m$ is the sectionally
holomorphic function $m_{\pm}$, is an entire function of
$\lambda$. 

Since $m_-\in G_3^-$,
$$
m_-=\sum_{j\ge 0}m_jJ_{\lambda}^{-j}, \qquad m_j\in \Cal E_j.
$$ 
In particular,  $m_-\sim I+\ell$, as $\lambda \rightarrow \infty$,
where $\ell$ is
strictly lower triangular. Since $\ell$ depends on $x$,
$(Dm_-)m_-^{-1}$ is bounded as $\lambda\rightarrow \infty$ while
$$
\align
m_-J_{\lambda}m_-^{-1}-J_{\lambda}=&[m_-,J_{\lambda}]m_-^{-1}\\
=&\left( \sum_{j\ge 0}(m_j-\sigma(m_j))J_{\lambda}^{1-j}\right) m_-^{-1}\\
=&\left( \sum_{j\ge 2}(m_j-\sigma(m_j))J_{\lambda}^{1-j}\right) m_-^{-1}.
\endalign
$$ 
The above expression is bounded as
$\lambda\rightarrow \infty$, and therefore so is
$$
(Dm_-)m_-^{-1}+m_-J_{\lambda}m_-^{-1}-J_{\lambda}.
$$
It follows by Liouville's theorem that this expression is independent of
$\lambda$. 

Let us call it $q(x,t)$. From the expansion for $m_-$ we conclude that
$q$ may be expanded as
$$  
q=\sum_{k=0}^{n-1}q_kJ_{\lambda}^{-k},
$$
where $ \quad q_k\in \Cal F_k$ since $q$ is independent of $\lambda$.

Writing (5.1.3) in the form  
$Dm+mJ_{\lambda}=(J_{\lambda}+q)m$ and expanding
$q$ and $m$ in powers of $J_{\lambda}$, we obtain a sequence
of recursion relations for $q_k$ (noting that $m_0=I$):
$$
Dm_k+(I-\sigma)(m_{k+1})=\sum_{j=0}^kq_j\sigma^{-j}(m_{k-j})=q_0m_k+\dots
+q_k,  \tag5.1.4
$$
for $0\le k \le n-1$. 

Since $m_-\in G_3$, $m_1=\omega_{10}I,$ and $q_0=(I-\sigma)(m_1)=0$.
From (5.1.4) we get for $k=1$ 
$$
\align
q_1=&m_1+(I-\sigma)(m_2)\\
=&D\omega_{10}I-(\sigma-I)(\omega_{20}I+\omega_{21}(\sigma-I)^{-1}
D_1\\
=&D\omega_{10}I-\omega_{21}D_1.\\
\endalign
$$ 
Since $q_1\in \Cal F_1$ its first entry vanishes, so
$\omega_{21}=D\omega_{10},$ and
$$
q_1=nD\omega_{10}e_{nn}.
$$ 

We now prove, by induction, that in fact $q_j$ is a scalar multiple
of $e_{nn}$
for all $1 \le j\le n-1$. Assume this is true for all integers less
than $j$. Since the
$m_s$ are diagonal matrices, $q_1m_{j-1}+\dots +q_{j-1}m_1$
on the right side of (5.1.4) is a scalar multiple of $e_{nn}$,
so we need to prove that
$$
Dm_j+(I-\sigma)(m_{j+1})=d_je_{nn}   \tag5.1.5
$$
for some constant $d_j$.
Since $q_j\in \Cal F_j$ the right side of 
(5.1.4) belongs to $\Cal F_j$, for $k=j$, and so
$$
\text{tr} \ e_{ss}[Dm_j+(I-\sigma)(m_{j+1})]=0, \quad s\le j.  \tag5.1.6
$$  
Writing
$$  
\gather
m_j=\omega_{j0}I+\sum_{k=1}^{j-1}\omega_{jk}(\sigma-I)^{-k}D_1,\\
m_{j+1}=\omega_{j+1,0}I+\sum_{k=1}^{j}\omega_{j+1,k}(\sigma-I)^{-k}D_1,
\endgather 
$$
we find, after some calculations, that (5.1.6) reduces to the
linear system of equations
$$ 
\text{tr} \ e_{ss}(D\omega_{j0}I-\omega_{j+1,1}D_1)+\sum_{k=1}^{j-1}
(D\omega_{jk}-\omega_{j+1,k+1})tr\ e_{ss}(\sigma-I)^{-k}D_1=0, \quad s\le j.
$$
This is the homogeneous system of equations corresponding to
(4.2.3); and as we saw in
the proof of Theorem 4.2.1, that system has a unique solution.
Therefore $D\omega_{j0}=\omega_{j+1,1}$,
$$ 
Dm_j+(I-\sigma)(m_{j+1})=nD\omega_{j0}e_{nn}
$$
and (5.1.5) is proved.
 
We now prove that $q$ satisfies the $(n,k)$ Gel'fand-Dikii equations. Since
$G_3^-$ is not a group, we cannot use the Lie algebra decomposition
that worked for the AKNS hierarchy. 
Instead, we argue as follows. The solution of the factorization problem
satisfies the simultaneous equations (3.2.9a,b). Therefore we have
$$
(D-\J-q)\psi=0, \qquad (D_t-G_k)\psi=0,  \tag5.1.7
$$
where $G_k=\J^k+B_{n,k}$ is the right side of (3.2.9b),
and $\psi=m\exp\{x\J+t\J^k\}$. From the first equation we
see that $\psi$ is a Wronskian: $\psi=||D^{j-1}v_{i-1}||,\ 
i,j=1,\dots,n$, and that the entries of the first row of
$\psi$ satisfy $Lv=\lambda v$. If $x_0$ is a regular point of
$q$, then these equations are also satisfied by
$\tilde \psi=\psi(x,\lambda)\psi^{-1}(x_0,\lambda)$, and $\tilde \psi(x_0,\lambda)
=I$. We shall henceforth assume this is the case and drop the tilde.

By the second equation each $v_j$
satisfies
$$
\pd{v}{t}=\sum_{j=1}^nG_{1j}D^{j-1}v
$$
where the $G_{1j}$ denote the entries of the first row
of $G_k$. Since $Lv_j=\lambda v_j$, we can replace 
$\lambda^s v$ by $L^sv$. The result is that each $v_j$ satisfies
an equation of the form $\dot v_j=P_kv_j$ where $P_k=D^k+\dots$
is a differential operator of order $k$ and is independent of 
$\lambda$. The simultaneous
equations $Lv_j=\lambda v_j$, and $\dot v_j=P_kv_j$ imply that
$(\dot L-[P_k,L])v_j=0;$
and moreover, $D^mv_j(x_0,z)=\delta_{jm}$ for $0\le j,m\le n-1$. 
 Let
$$
[P_k,L]=\sum_{j=0}^kr_j(x)D^j.
$$
Then
$$
0=\left( \dot L -[P_k,L]\right) v_j(x_0,z)=\dot u_j(x_0)-r_j(x_0)
-\sum_{m=n}^kr_m(x_0)D^mv_j(x_0,z),
$$
where the sum over $m\ge n$ is vacuous if $k<n$. Since
derivatives of $v$ of order $n$ or greater can be replaced
by $\lambda$, the summation is a polynomial in $\lambda$ of degree
at least one. Since the above equation holds identically in
$\lambda$, we conclude that the summation vanishes identically,
$[P_k,L]$ is of order less
than $n$, and $\dot L-[P_k,L]$ vanishes at $x_0$. Since $x_0$ was arbitrary,
it vanishes at all regular points of $q$.

This implies that $[P_k,L]$ is
of order $n-2$. It follows  (cf. Proposition 2.3 [DS]) that
the coefficients of $P_k$ are differential polynomials
of the coefficients of $L$ and that $P_k$ is a linear combination
of the operators  $L^{j/n}_+$ for $j\le k$:
$$
P_k=\sum_{j=1,\ j\ne \ell n}^k c_j(L^{j/n})_+,
$$
for some constants $c_j$.

To show that $P_k$ is precisely $L^{k/n}_+$, it suffices
to prove that all the $c_j,\ j<k$, are zero. We do this by showing
that for $j<k$
 the coefficient of $D^j$ in $P_k$ is a differential polynomial in
the entries of $q$ with zero constant term. The coefficients
of $P_k$ come from the first row of $G_k$, and this matrix is
in turn given by (3.2.9b). From the lower triangular structure
of the leading term of $m\in G_3^-$, the first row of $G_k$ is 
identical to that of $P_1(m\J^k m^{-1})$. But the entries
of $m\J^km^{-1} $ are differential polynomials in $q$ 
with zero constant terms by Theorem 3.2.6. \qed

\enddemo

\subhead 5.2 The Modified Gel'fand-Dikii flows \endsubhead
 The so-called modified Gel'fand-Dikii flows are a 
special case of the AKNS hierarchy, and are defined as
follows. Writing $L=(D+u_n)(D+u_{n-1})\cdots (D+u_1)$, the
equation $Lv=\lambda v$ can be written as a first order system
$$
(D-\J+q)\psi=0
$$
where $q$ is the diagonal matrix with entries $u_1,\dots,u_n$.
( $\tr\, q=0$ since the coefficient of $D^{n-1}$ vanishes.)

Recall (\S2.1) that $\Lambda_z^{-1}\J \Lambda_z=zJ_{\alpha}$,
where $J_{\alpha}$ is diagonal. Since $q$ is diagonal,
 $\Lambda_z^{-1}q \Lambda_z=\tilde q$ is independent of $z$.
Hence we can write the isospectral problem as
$$
(D-zJ_{\alpha}+\tilde q)\tilde \psi=0.
$$
This is the isospectral problem for an AKNS system, whose flows
we constructed in \S3.2.

Working in the original basis, we obtain the $(n,k)$
modified Gel'fand-Dikii flow as
$[D_x,D_t]=0$, where
$$
D_x=\pd{}{x}-\J+ q,
\qquad
D_t=\pd{}{t}-[F^k]_+,
$$
$F=m\J m^{-1}$, and $[\cdot]_+$ denotes the projection $P_2$,
i.e. the projection onto non-negative powers of $\J$.

The modified Gel'fand
Dikii flows are obtained from the second factorization:

\proclaim{5.2.1 Theorem} Let $g\in G$ and let $m_{\pm}$ be the holomorphic
solutions to the factorization problem
$$
e^{i(xJ_{\lambda}+tJ_{\lambda}^k)}ge^{-i(xJ_{\lambda}+tJ_{\lambda}^k)}=
m_-^{-1}m_+, \qquad m_{\pm}\in G_2^{\pm}.
$$
Then 
$$
q(x,t)=(Dm)m^{-1}+mJ_{\lambda}m^{-1}-J_{\lambda}
$$
is a diagonal matrix of trace zero and satisfies the modified
Gel'fand-Dikii equations for (k,n).
\endproclaim

\demo{Proof} We repeat the first part of the proof of Theorem 5.1.1.
Again we may conclude that 
$$
(Dm)m^{-1}+mJ_{\lambda}m^{-1}-J_{\lambda}  \tag5.2.2
$$
 is an entire function of
$\lambda$, where $m$ is the sectionally meromorphic function
defined by $m_+,\ m_-$ in the regions $|\lambda|<\rho$, 
$|\lambda|>\rho$ respectively.
This time
$$
m_-=\sum_{j\ge 0}m_jJ_{\lambda}^{-j}, \quad m_0=I,\ m_j\in \Cal D. \tag5.2.3
$$
Again $m_-=I+\ell +O(\lambda^{-1})$ as
$\lambda \rightarrow \infty$ and so  
the expression in (5.2.2) is bounded as
$\lambda$ tends to infinity. As before, 
we denote it by $q$, where $q$
depends only on $x$ and $t$. Inserting (5.2.3) into (5.2.2), we obtain
$q$ as the constant term in
$$ 
\gather
m_-J_{\lambda}m_-^{-1} \\
=(I+m_1J_{\lambda}^{-1}+\dots)J_{\lambda}(I-m_1J_{\lambda}^{-1}+\dots)-J_{\lambda} \\
=m_1-J_{\lambda}m_1J_{\lambda}^{-1}=(I-\sigma)(m_1)+\dots
\endgather
$$ 
where the dots denote negative powers of $J_{\lambda}$. Thus
$$
 q=(I-\sigma)(m_1)\in \Cal D_0.
$$

To show that $q$ satisfies the modified Gel'fand-Dikii 
$(n,k)$ flow, we consider the connection
$D-J_{\lambda}-q, \ D_t-v_k$, where $v_k$ is given by the right side of
(3.2.9b). 
 As in the proof of Theorem 3.2.6, this connection is flat.
This time $G_2^-$ is a group, so we obtain the Lie-algebra
decomposition  
$$
\align
v_k=&P_2\left[ (D_t \tilde m)\tilde m^{-1}+ 
\tilde mJ_{\lambda}^k\tilde m^{-1}\right] \\
= &P_2(\tilde mJ_{\lambda}^k\tilde m^{-1})\\
=&\sum_{j=0}^kF_j\J^{k-j}.
\endalign
$$
By conjugating by the constant matrix $\Lambda_{\alpha}$
everything is expanded in powers of $zJ_{\alpha}$,
($F_j$ is diagonal, so it commutes with $d(z)$)
and we see that this is precisely the $k^{th}$ AKNS flow.
\qed \enddemo
                                                       
\subhead 5.3 The ``potential" Gel'fand-Dikii flows \endsubhead
Factorizations in $G_1^{\pm}$ lead to a third
structure for the potential. This time $m\in G_1^-$ and so has the expansion
$$
m=\sum_{j=0}^{\infty}\frac{m_j}{\lambda^j},\quad m_0=I.
$$
The arguments of the previous section are repeated, this time using
the projection 
$P_1$. From (5.2.2) we obtain
$$  
P_1( (Dm)m^{-1}+mJ_{\lambda}m^{-1} ) 
                      =P_1(mJ_{\lambda}m^{-1})
                      =J_{\lambda}+[m_1,e_{n1}] 
$$
and one finds that
$$
q= \left( \matrix m_{1n} & 0 & \hdots & 0 \\
                                        m_{2n} & 0 & \hdots  & 0\\
                                        & \vdots   &  & \\
                                        m_{nn}-m_{11}& -m_{12}& \hdots & -m_{1n}
                        \endmatrix \right),  
$$

where $m_{jk}$ are the entries of the coefficient matrix $m_1$. The dressing argument goes through as in the previous section. The following is related to the general theory of Drinfeld and Sokolov. As they show, all
reductions of the $n^{th}$ order equation $Lv=
\lambda v$ to a first order system are related by a gauge transformation,
and the entries of $L$  are differential polynomials
in the coefficients the potential. 
\proclaim{5.3.1 Theorem}  The entries of $q$ are differential
polynomials of precisely $n-1$ independent functions.
\endproclaim

We first need to prove the following lemma.
\proclaim{5.3.2 Lemma} For $m\in G_1^-$ we have the representation
$$
m=\sum_{j=0}^{\infty}m_jJ_{\lambda}^{-j}, \qquad m_0=I, \ m_j\in \Cal D  \tag5.3.3
$$
where the diagonal matrices $m_1, \dots, m_n$ satisfy the additional constraints
$$
\gather
\tr\, m_je_{kk}=0 \quad  j=1,\dots , n-1;\ k\ge j+1 \tag5.3.4 \\
\tr\, m_j=0\ \text{for}\ j=0\ \text{mod}\ n. \tag5.3.5
\endgather
$$
\endproclaim
\demo{Proof} From the form of $J_{\lambda}^{-1}$, namely
$$
J_{\lambda}^{-1}=\left( \matrix 0 &\hdots & & \lambda^{-1} \\
                                 1 & 0 & \hdots & \\
                                0 & 1 & 0 & \hdots\\
                                  &   & \ddots & \\
                                      & & 1 & 0         \endmatrix \right),
$$ 
we see that the constraints (5.3.4) are necessary and sufficient that
the constant term in $m$ be the identity matrix. The constraint $tr\ m_n=0$
is a consequence of the condition $\det \ m =1$. \qed \enddemo
  
Now write $q$ in the form
$$
q=\sum_{j=0}^{n-1}q_jJ_{\lambda}^{-j}
$$
where
$$
q_0=u_0(e_{11}-e_{nn}), \quad q_j=u_j(e_{jj}-e_{nn})+v_j(e_{jj}+e_{nn}),
\quad q_{n-1}=v_{n-1}e_{nn}.
$$ 
Substituting this form of $q$, together with the representation (5.3.3)
into the equation  $Dm=[J_{\lambda},\ m]+qm$, we obtain the recursion relations
$$
\gather
m_1-\sigma(m_1)=q_0,  \\
Dm_1+m_2-\sigma(m_2)=q_0m_1+q_1, \\
\vdots \\
Dm_j+m_{j+1}-\sigma (m_{j+1})=\sum_{k=0}^{n-1}q_k\sigma^{-k}(m_{j-k}).
\endgather
$$  

We solve these equations recursively. From the first  we get $m_1=u_0e_{11}$
and $q_0=u_0(e_{11}-e_{nn})$. Taking the trace of the second equation
we find $tr\ q_1=2v_1=Du_0-u_0^2$. With this choice of $v_1$ the 
diagonal matrix $m_2$ is uniquely determined by the second equation and
the constraint (5.3.4). This process continues up to $j=n-2$. At each
stage the parameters $v_j$ are determined in terms of the preceeding
$u_0, \dots,\ u_{j-1}$, and the $m_{j+1}$ is uniquely determined.

The equation at $j=n-1$ is
$$
Dm_{n-1}+m_n-\sigma (m_n)=q_0m_{n-1}+\dots+q_{n-1}.  \tag5.3.5
$$
The first $n-1$ rows of (5.3.5)
plus the trace condition $tr\ m_n=0$ uniquely determine the entries
of $m_n$ as differential polynomials in $u_j,\ v_j$ for $j=0, \dots, n-2$
The last row of (5.3.5) then determines $v_{n-1}$

We thus see that the entries of $q$ are differential polynomials of
the functions $u_0,\dots, u_{n-1}$. \qed

For $n=2,3$ we get
$$
q=\left( \matrix u & 0\\ Du-u^2 & -u \endmatrix \right),
\qquad
\left( \matrix u_0 & 0 & 0 \\ \frac{u_1+(D-u_0)u_0}{2} & 0 & 0 \\
(D-u_0)u_1 & \frac{-u_1+(D-u_0)u_0}{2}& -u_0 \endmatrix \right).
$$

This class of potentials leads to a third class of flows, the
``potential Gel'fand-Dikii" flows. These flows may be obtained
as a reduction of the potential Kadomtsev-Petviashvili hierarchy [Sz].


\Refs 

\ref \no {B}\by {R. Beals} 
\paper The inverse problem for ordinary differential operators on
the line
\jour American Journal of Mathematics
\yr 1985 \vol 107 \pages 281-366
\endref

\ref \no {BDT}
 \by {R. Beals, P. Deift, and C. Tomei} 
\book Direct and Inverse Scattering on the Line 
\publ Amer. Math. Soc. \publaddr Providence, R.I.\yr 1989
\endref
                              

\ref \no {BS1}  \by {R. Beals, and D.H. Sattinger}   
\paper Action-angle variables for the Gel'fand-Dikii flows 
\yr 1992 \vol 43
\jour Zeitschrift f\"ur Angewandte Mathematik und Physik, 
 \endref

\ref \no {BS2} \bysame
\paper Integrable Systems and Isomonodromy Deformations
\yr 1992
\jour preprint
\endref

 
\ref \no {D} \by {J. Dorfmeister}
\paper Banach manifolds of solutions to nonlinear partial differential
equations, and relations with finite dimensional manifolds
\inbook Differential Geometry \eds R.E.Greene, S.-T.Yau 
\publ American Math Society
\endref      

\ref \no DZ \by {P. Deift and X. Zhou}  \pages 485-533
\paper Inverse scattering for nth order differential operators
\yr 1991 \vol 54        
\jour Communications of Pure and Applied Mathematics 
\endref

\ref \no DS \by {V.G. Drinfel'd and V.V. Sokolov} \pages 1975-2036
\paper Lie algebras and equations of Korteweg-deVries type
\yr 1985  \vol 30
\jour  Journal of Soviet Mathematics  
\endref

 \ref \no GD  \by {I.M. Gelfand, and L.A. Dikii} \pages 259-273
\paper Fractional powers of operators and Hamiltonian systems 
\yr 1976 \vol 10
\jour Functional Analysis and its Applications.
\endref 
 
\ref \no GK 
\by {I.C. Gohberg and M.G. Krein} 
\pages 217-284 
\paper Systems of integral equations on a half line with kernels depending on the difference of arguments
\transl American Math. Soc. Transl. 
\yr 1960 \vol 14 \pages 217-287
\endref

\ref \no K \by {V. Kac}
\book Infinite dimensional Lie algebras
\publ  Cambridge University Press
\publaddr Cambridge, U.K. \yr  1985
\endref


\ref \no KW \by {B.A. Kuperschmidt and G. Wilson}  
\paper Modifying Lax equations and the second Hamiltonian structure
\jour Invent. Math. 
\yr 1981 \vol 62 \pages 403-436
\endref   

\ref \no Mal \by {B. Malgrange}
\book La Classification des connexions irregulieres a une
variable
\publ Universit\'e de Grenoble \publaddr Grenoble, France \yr 1982
\endref 

\ref \no M \by {T. Miwa}
\paper Painlev\'e property of monodromy preserving equations and the
analyticity of the $\tau$ function
\jour Publ. R.I.M.S. Kyoto 
\yr 1981, \vol 17  \pages 703-721
\endref 

\ref \no PS  \by {A. Pressley and G. Segal}
\book Loop groups
\publ Oxford University Press \publaddr Oxford, U.K. \yr 1986
\endref



\ref \no Sch \by {R. Schilling}
\paper A loop algebra decomposition for Korteweg-deVries
equations
\inbook Integrable and Superintegrable Systems  \eds B. Kupershmidt
\publ World Scientific Press \yr 1990
\publaddr Singapore
\endref



\ref \no SW \by {G. Segal and G. Wilson}
\paper Loop groups and equations of KdV type
\jour Publications I.H.E.S. 
\yr 1985 \vol 61
\endref 

\newpage


\ref \no Si \by {B. Simon}  \pages Cambridge University Press
\book Trace ideals and their applications
 \publ London Mathematical Society Lecture Note Series
\yr 1979 \vol 35
\endref
           

 \ref \no Sz \by {J. Szmigielski} 
 \book Infinite dimensional homogeneous manifolds
with translational symmetry and nonlinear
partial differential equations
\bookinfo Disseration, University of Georgia-Athens, Ga. 
\yr 1987
  \endref


 

\ref \no ZS
 \by {V.E. Zakharov and A. Shabat}
\paper Integration of nonlinear equations of
mathematical physics by the method of the inverse scattering, II
\jour Funktsional'nyi Analiz i Ego Prilozheniay
 \yr 1979  \vol 13 \pages 13-22
\transl \jour Functional Analysis and its Applications
\endref

\endRefs
\enddocument